\documentclass[fleqn,usenatbib]{mnras}

\usepackage[T1]{fontenc}
\usepackage{ae,aecompl}
\usepackage{amsmath,amssymb}
\usepackage{graphicx}
\usepackage{longtable,lscape}
\usepackage{caption}
\usepackage{color}
\usepackage{ulem}
\usepackage{threeparttable}
\usepackage{comment}
\usepackage{hyperref}
\hypersetup{
    colorlinks=true,
    linkcolor=blue,
    citecolor=blue,
    urlcolor=blue
}
\usepackage{adjustbox} 
\usepackage{subcaption} 
\usepackage{caption} 

\usepackage{orcidlink} 
\usepackage{txfonts}   

\newcommand{\email}[1]{\href{mailto:#1}{\texttt{#1}}}

\title[A possible origin of the overlapping light curve of eRO-QPE1]{A possible origin of the overlapping light curve of eRO-QPE1}

\author[Xu et al.]{
Wen-Long Xu\orcidlink{0009-0003-9792-9325},$^{1}$ 
Hui-Hui Wang\orcidlink{0000-0003-3321-9458},$^{2}$ 
Yi-Gu Chen\orcidlink{0000-0002-8043-6650},$^{1}$ 
Yu-Zhe Li\orcidlink{0009-0002-9215-5618},$^{1}$ 
Wei-Hua Lei\orcidlink{0000-0003-3440-1526}$^{1}$ \\ 
$^{1}$ School of Physics, Huazhong University of Science and Technology, Wuhan 430074, China; \email{leiwh@hust.edu.cn} \\
$^{2}$ School of Physics and Engineering, Henan University of Science and Technology, Luoyang 471023, China;\email{wanghh0201@haust.edu.cn}\\
}

\begin{document}
\maketitle

\begin{abstract}
Quasi-Periodic Eruptions (QPEs) are recurrent X-ray eruptions found so far in the nuclei of low-mass galaxies. Their trigger mechanism is still unknown, and new observations are demanded to reveal the physics behind it. Recently, chaotic mixtures of multiple overlapping eruptions with varying amplitudes have been observed in eRO-QPE1 obs1. This complex behavior presents a challenge to the existing QPE models. In this paper, we propose that the overlapping features may be the result of gravitational lensing. Therefore, all QPEs could share a common trigger mechanism, and we do not need a special mechanism for eRO-QPE1.
\end{abstract}


\section{Introduction}
\label{intro}

Quasi-Periodic Eruptions (QPEs) are a recently discovered class of astrophysical transients, characterized by high-amplitude eruptions and regular flares in soft X-ray band, typically within the $0.5$-$3$ keV range. 
These eruptions involve rapid and repetitive increases in the X-ray count rate—often by more than an order of magnitude above a stable quiescent level \citep{2020A&A...636L...2G}. To date, approximately 11 QPEs have been detected \citep{2019Natur.573..381M,2020A&A...636L...2G,2021Natur.592..704A,2021ApJ...921L..40C,2023A&A...675A.152Q,2023NatAs...7.1368E,2024A&A...684A..64A,2024Natur.634..804N,2025NatAs.tmp...99H}. 

The known QPEs share common observational traits: they recur quasi-periodically, exhibit short durations (from tens of minutes to several hours) compared to their recurrence intervals (hours to days), and are typically found in the nuclei of low-mass galaxies hosting central black holes with masses ranging from $\sim 10^5$ to $10^7 M_\odot$ \citep{2019Natur.573..381M, 2021Natur.592..704A, 2024Natur.634..804N}. 
The first identified source, GSN 069, emerged from a long-lived tidal disruption event \citep{2011arXiv1106.3507S, 2013MNRAS.433.1764M, 2018ApJ...857L..16S} and shows $\sim$1-hour eruptions recurring every $\sim$ 9 hours \citep{2019Natur.573..381M}. 
Subsequent discoveries revealed a diverse population. 
RX J1301.9+2747 showed flares lasting about half an hour with variable recurrence times \citep{2013ApJ...768..167S, 2020A&A...636L...2G}. 
AT 2019vcb was first identified as an optical TDE before revealing QPEs \citep{2023A&A...675A.152Q}, and XMMSL1 J024916.6-041244 was found via archival search \citep{2021ApJ...921L..40C}. 
Systematic searches have uncovered several more, including eRO-QPE1 and eRO-QPE2 \citep{2021Natur.592..704A}, eRO-QPE3 \citep{2024A&A...684A..64A}, eRO-QPE4 \citep{2024A&A...684A..64A} and eRO-QPE5 \citep{2025ApJ...989...13A}, AT2019qiz \citep{2024Natur.634..804N}, and SwJ023017.0+283603 which exhibits outbursts on weekly timescales \citep{2023NatAs...7.1368E}. 
This diversity is exemplified by their varying properties: eRO-QPE1 has rise-decay durations of $\sim$ 7.6 hours and separations of $\sim$ 18.5 hours, while eRO-QPE2 shows much shorter, $\sim$ 27-minute eruptions recurring every $\sim$ 2.4 hours \citep{2021Natur.592..704A}. 
The most recent source, ansky, displays extreme properties with a $\sim$ 4.5-day period and a $\sim$ 25-day superperiod \citep{2025NatAs.tmp...99H}. 
Recent follow-up on ansky indicates its QPEs are evolving, becoming more energetic with longer recurrence times and more asymmetric profiles \citep{2025A&A...703A.263H}, and its UV properties favor a TDE origin over an AGN \citep{2025ApJ...994L..16Z}. 
Low-significance optical/UV counterparts have also been reported for some QPEs \citep{2024A&A...688A.157S, 2025NatAs.tmp...99H}.

However, despite extensive observational efforts, the origin of QPEs remains uncertain. 
A variety of models have been proposed to explain the origin of QPEs, including accretion disk instabilities \citep{2019Natur.573..381M,2020A&A...641A.167S,2021ApJ...909...82R,2022ApJ...928L..18P,2023ApJ...952...32P}, gravitational lensing in a black hole binary with a mass ratio close to unity \citep{2021MNRAS.503.1703I}, star-disk interactions \citep{2021ApJ...921L..32X,2023A&A...675A.100F,2023ApJ...957...34L,2024PhRvD.109j3031Z,2024PhRvD.110h3019Z,2025ApJ...978...91Y,2024ApJ...963L...1L,2025arXiv250502596X} and QPEs from stable/unstable mass transfer due to Roche lobe overflow from a main-sequence star have also been proposed \citep{2023MNRAS.524.6247L,2023ApJ...945...86L,2024A&A...687A.295W}. 
Models involving a two- or three-body system comprising of a massive black hole and one or more stellar-mass companion, have also attracted significant attention \citep{2022ApJ...933..225W,2020MNRAS.493L.120K,2022A&A...661A..55Z,2021ApJ...921L..32X,2022ApJ...926..101M,2023ApJ...947...32C,2025ApJ...983L..18J,2025MNRAS.543.3503Y}. 
Such systems could potentially emit gravitational wave signals detectable by future observatories like LISA and TianQin \citep{2007CQGra..24R.113A,2017PhRvD..95j3012B,2022A&A...661A..55Z,2022ApJ...930..122C}. 
The possible connection between QPEs and TDEs is also under active debate \citep{2021ApJ...921L..40C,2023ApJ...957...34L,2023A&A...670A..93M,2023A&A...675A.152Q,2024arXiv240916908B, 2024Natur.634..804N,2024A&A...682L..14W,2024arXiv240910486G,2024ApJ...970L..23W,2025arXiv250319722X,2026arXiv260113113W}. 

A particularly intriguing case is eRO-QPE1, which exhibits a chaotic mixture of multiple overlapping eruptions with markedly different amplitudes \citep{2022A&A...662A..49A,2024ApJ...965...12C}. 
This overlapping pattern in the light curve may suggest that, beyond examining the overall evolutionary trends of QPEs, it is also important to consider the presence and nature of substructures within individual events.  
Such complexity challenges current QPE models and raises the prospect that an external modulation—rather than a change in the intrinsic QPE mechanism—may be shaping the observed light‑curve morphology. One proposed explanation for this complexity is that two stellar extreme-mass-ratio inspirals (EMRIs) could undergo Roche-lobe overflow at periapse, producing the observed overlapping eruptions \citep{2022A&A...662A..49A, 2022ApJ...926..101M}.

In this paper, we propose that the complex overlapping eruptions observed in eRO-QPE1 may be explained by gravitational lensing. 
Under this interpretation, the source could potentially be accommodated within existing QPE frameworks, without the need to invoke a separate or exotic mechanism. 
According to gravitational lensing theory, photons from a background source can be deflected by a lensing object, resulting in multiple images or producing a local magnification of the portion of the light curve that lies within the Einstein radius \citep{schneider1992gravitational}. 
These images typically have different magnifications but preserve the same temporal profiles, exhibiting time delays and amplitude variations, while their spectral shapes share the same trend \citep{schneider1992gravitational}. 
Initially, by analogy with \cite{2021NatAs...5..560P}, we aimed to analyze as systematically as possible whether the eRO‑QPE1 Obs1 light curve exhibits evidence of lensing features. 
Unfortunately, owing to discrepancies between strong lens model and the observational data, we proceeded to consider a microlensing model, which we briefly describe in the Section \ref{sec:summary}.

This paper is organized as follows. Section \ref{sec:theory} introduces the basic theory of gravitational lensing, including the PM model for point-mass lenses such as Schwarzschild black holes, and the singular isothermal sphere lens model for galaxies with specific mass distributions. Section \ref{sec: data analysis} presents evidence for gravitational lensing in eRO-QPE1 based on data analysis and light curve fitting. 
Section \ref{sec:optical_depth} discusses the optical depth. Finally, we summarize our results and discuss the microlensing scenario in Section \ref{sec:summary}.

\section{Basic theory of gravitational lensing}
\label{sec:theory}
In this section, we briefly introduce two lens models: the point mass (PM) lens model and the singular isothermal sphere (SIS) lens model \citep{schneider1992gravitational}.  Considering a PM lens model, the light will be deflected with an angle of $\alpha$\ in the limit of geometric optics.
\begin{eqnarray}
\alpha_{\pm}=\frac{4GM_{\rm Lz}}{c^2 \xi_{\pm}}.
\label{eq1}
\end{eqnarray}
where $M_{\rm Lz}$ is the lens mass, $\xi_{\pm}$ is the impact parameter denoting the closest distance between the light ray and the lens in the lens plane, and $G$ and $c$ are the gravitational constant and the speed of light, respectively.

\begin{figure}
    \centering
    \includegraphics[width=\columnwidth]{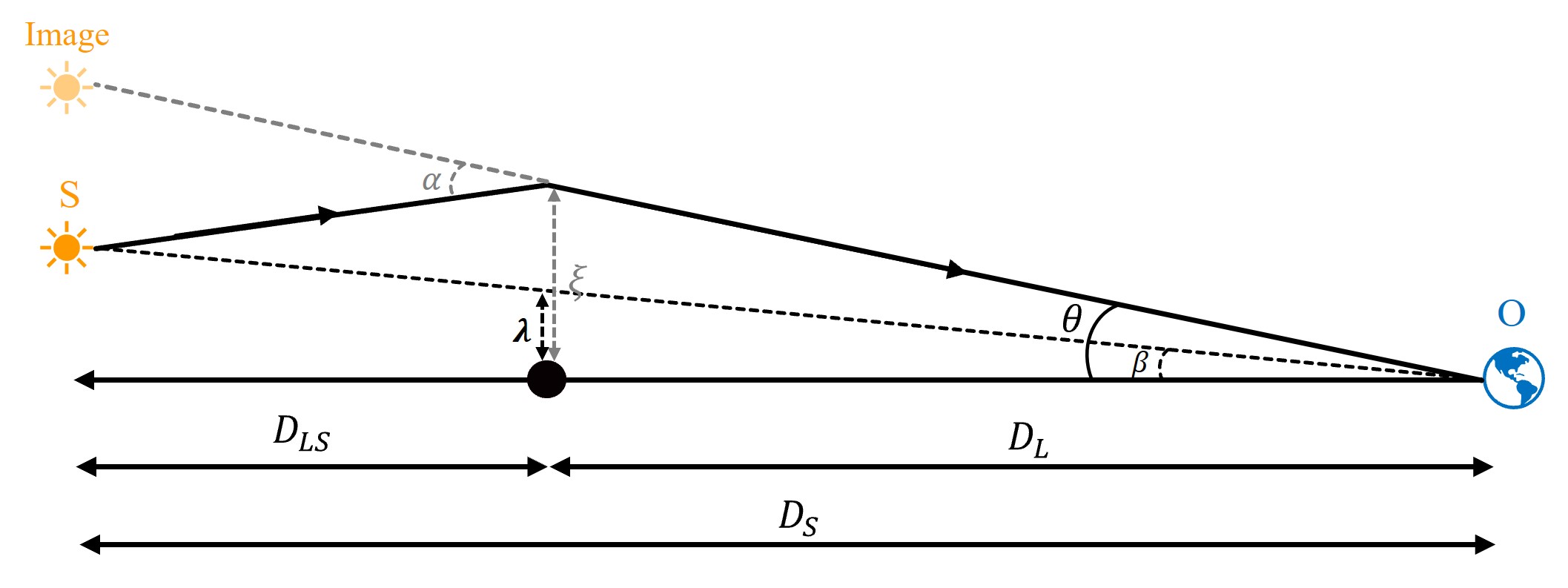}
    \caption{The geometry of point mass lensing system. The source is marked as $\rm{S}$, and the observer is located at $\rm{O}$. $D_{\rm S}$ and $D_{\rm L}$ are the distance from the source to the observer and the lens object to the observer, respectively. $D_{\rm LS}$ is the separation between source and lens object. $\alpha, \beta, \theta$ and $\xi$ represents the deflection angle, angular position of the source without lens, image position and impact parameter, respectively. } 
    \label{fig:lenssystem}
\end{figure}

The basic geometric configuration of the point mass lens model is shown in Figure \ref{fig:lenssystem}. Wherein $D_{\rm LS}$, $D_{\rm S}$, and $D_{\rm L}$ label the lens-source distance, the source-observer distance, and the lens-observer distance, respectively. The angular separation of the image and the source is naturally expressed as $\theta$ and $\beta$. Based on the small angle approximation, we get the lens equation.
\begin{eqnarray}
D_{\rm LS}\alpha+D_{\rm S}\beta=D_{\rm S}\theta,
\label{eq2}
\end{eqnarray}
Based on the geometry of $\xi_{\pm} \approx \theta_{\pm} D_{\rm L}$. One can solve the lens equation and find two solutions.
\begin{eqnarray}
\theta_{\pm}=\frac{1}{2}[\beta \pm (\beta^2 + \frac{16GM}{c^2} \frac{D_{\rm LS}}{D_{\rm L} D_{\rm S}})^{1/2}],
\label{eq3}
\end{eqnarray}
where "$+$" and "$-$" denote the parity of the image. 
For small angles, one can then solve for $\xi$ by multiplying both sides of Equation. (\ref{eq3}) with $D_{\rm L}$ to obtain \citep{1991ApJ...378...22K} 
\begin{eqnarray}
\xi_{\pm}=\frac{1}{2} [\lambda \pm \sqrt{\lambda^2 +8R_{\rm S}\frac{D_{\rm LS}D_{\rm L}}{D_{\rm S}}}],
\label{eq4}
\end{eqnarray}
and $\lambda=D_{\rm L} \beta$ corresponds to the distance from the light ray to the lens object in the lens plane when the lens object is neglected. $R_{\rm S} = 2GM_{\rm Lz}/c^2$ corresponds to the Schwarzschild radius of the lens. 
The lens equation involves an effective angle $\theta_{\rm E}$, given by
\begin{eqnarray}
\theta_{\rm E}=\sqrt{2R_{\rm S}\frac{D_{\rm LS}}{D_{\rm L}D_{\rm S}}}.
\label{eq5}
\end{eqnarray}
The Einstein radius is thus given as
\begin{eqnarray}
r_{\rm E} = D_{\rm L} \theta_{\rm E} = \sqrt{2R_{\rm S} \frac{D_{\rm L} D_{\rm LS}}{D_{\rm S}}}.
\label{eq6}
\end{eqnarray}
The magnification effect appears inside the Einstein radius \citep{1984ApJ...284....1T}. 
By defining a dimensionless parameter $y=\lambda / r_{\rm E}$ and $y_{\theta_{\pm}} = \xi_{\pm}/r_{\rm E}$, Equation. (\ref{eq4}) has the form of 
\begin{eqnarray}
y_{\theta_{\pm}}=\frac{1}{2}[y \pm \sqrt{y^2 +4}].
\label{eq7}
\end{eqnarray}
Since gravitational lensing changes the apparent solid angle of a source. The total flux received from a gravitationally lensed image of a source is therefore changed in proportion to the ratio between the solid angles of the image and the source, we can obtain the magnification of the two images \citep{1996astro.ph..6001N} 
\begin{eqnarray}
\mu_{\pm}= \frac{1}{4}\left[\frac{y}{\sqrt{y^2+ 4}}+\frac{\sqrt{y^2+4}}{y}\pm 2 \right],
\label{eq8}
\end{eqnarray}
where $\mu_+$ and $\mu_-$ denote the magnification of the positive  and negative image, respectively. Thus, the flux ratio for resolved images can be expressed as
\begin{eqnarray}
\mu=\frac{I_{\xi_+}}{I_{\xi_-}}=\frac{\mu_{+}}{\mu_{-}}=\frac{y^2 +2 +y \sqrt{y^2+4}}{y^2 +2 -y \sqrt{y^2+4}},
\label{eq9}
\end{eqnarray}
where $I_{\xi_+}$ and $I_{\xi_-}$ are the brightness of the positive image and the negative image, respectively.

The time delay arises from two aspects. On the one hand, it is caused by the geometric time delay due to the different paths taken by two light rays reaching the observer. On the other hand, it is caused by the Shapiro delay resulting from the two light rays passing through different gravitational potentials when they pass through the lens plane \citep{1964PhRvL..13..789S,weinberg1972gravitation}.
The time delay between these two images is
\begin{eqnarray}
\Delta t= \frac{D_{\rm L} D_{\rm LS}}{2cD_{\rm S}}(\alpha_-^2-\alpha_+^2)+\frac{2GM_{\rm Lz}}{c^3} \ln{(\frac{\xi_+^2}{\xi_-^2})}.
\label{eq111}
\end{eqnarray}
Combining equations above, one can rewrite the time delay as
\begin{eqnarray}
\Delta t= \frac{2GM_{\rm Lz}}{c^3}f(y),
\label{eq11}
\end{eqnarray}
where $M_{\rm Lz}=M_{\rm L} (1+z_{\rm L})$ is the redshift mass of the lens and $f(y)= y\sqrt{y^2 + 4}+\ln{\{[(y^2+2)+y\sqrt{y^2+4}]/[(y^2+2)-y\sqrt{y^2+4}]\}}$. 
By invoking Equation. (\ref{eq9}) and (\ref{eq11}), one can estimate the redshifted lens mass \citep{1991ApJ...378...22K,1992ApJ...389L..41M,1992ApJ...399..368N} 
\begin{eqnarray}
M_{\rm Lz}=\frac{c^3 \Delta t}{2G}(\frac{\mu-1}{\sqrt{\mu}}+\ln{\mu})^{-1}.
\label{eq12}
\end{eqnarray}

The SIS model is often used to describe a galaxy acting as the gravitational lens \citep{schneider1992gravitational,1994A&A...284..285K,1996astro.ph..6001N,2022MNRAS.516..453G}. The Einstein angle is defined as
\begin{eqnarray}
\theta_{\rm E} = \sqrt{\frac{4GM(\theta _ {\rm E})}{c^2} \frac{D_{\rm LS}}{D_{\rm L}D_{\rm S}}}=4\pi\frac{\sigma^2_v}{c^2}\frac{D_{\rm LS}}{D_{\rm S}},
\label{eq13}
\end{eqnarray}
where $M(\theta_{\rm E})$ is the lens mass within the Einstein radius and $\sigma_{\rm v}$ is the velocity dispersion of the galaxy. The corresponding lens equation can be defined in terms of $y^{\prime}=\beta/\theta_{\rm E}$ and $x=\theta/\theta_{\rm E}$ as
\begin{eqnarray}
y^{\prime} = x-\frac{x}{|x|},
\label{eq14}
\end{eqnarray}
when $y^{\prime}<1$, there are two solutions: $x_{\pm}= y \pm 1$. However, when $y^{\prime} > 1$, the gravitational effect is weak and the lens equation has only one solution: $x = y^{\prime}+1$, hard to justify the existence of the lens. Here, we focus on the case of $y^{\prime} < 1$. 
In this case, the time delay between different images reads
\begin{eqnarray}
\Delta t = \frac{32\pi^2}{c} (\frac{\sigma_{\rm v}}{c})^2 \frac{D_{\rm L} D_{\rm LS}}{D_{\rm S}}(1+z_{\rm L})y^{\prime},
\label{eq15}
\end{eqnarray}
and the magnifications of the images are
\begin{eqnarray}
\mu_{\pm} = |1\pm \frac{1}{y^{\prime}}|.
\label{eq16}
\end{eqnarray}
Combining Equation. (\ref{eq15}) and Equation. (\ref{eq16}), we can get a formula for lens mass estimation of the SIS model

\begin{eqnarray}
M_{\rm Lz} = \frac{c^3}{8G} \frac{\mu+1}{\mu-1} \Delta t,
\label{eq17}
\end{eqnarray}
where $\mu$ is the magnification ratio. 
So that we can estimate the redshifted lens mass once the magnification ratio and time delay has been observed.

\section{Clues of gravitational lensing}
\label{sec: data analysis}
In this section, we first fit the light curve using a lensing model. 
Based on the fitting results, we then perform a spectral analysis of the two sub-eruptions across the five energy bands. 
Finally, the mass of the lensing object is estimated from the posterior distribution of the fitting parameters. 

\begin{figure}  
\centering
\includegraphics[width=8.2cm]{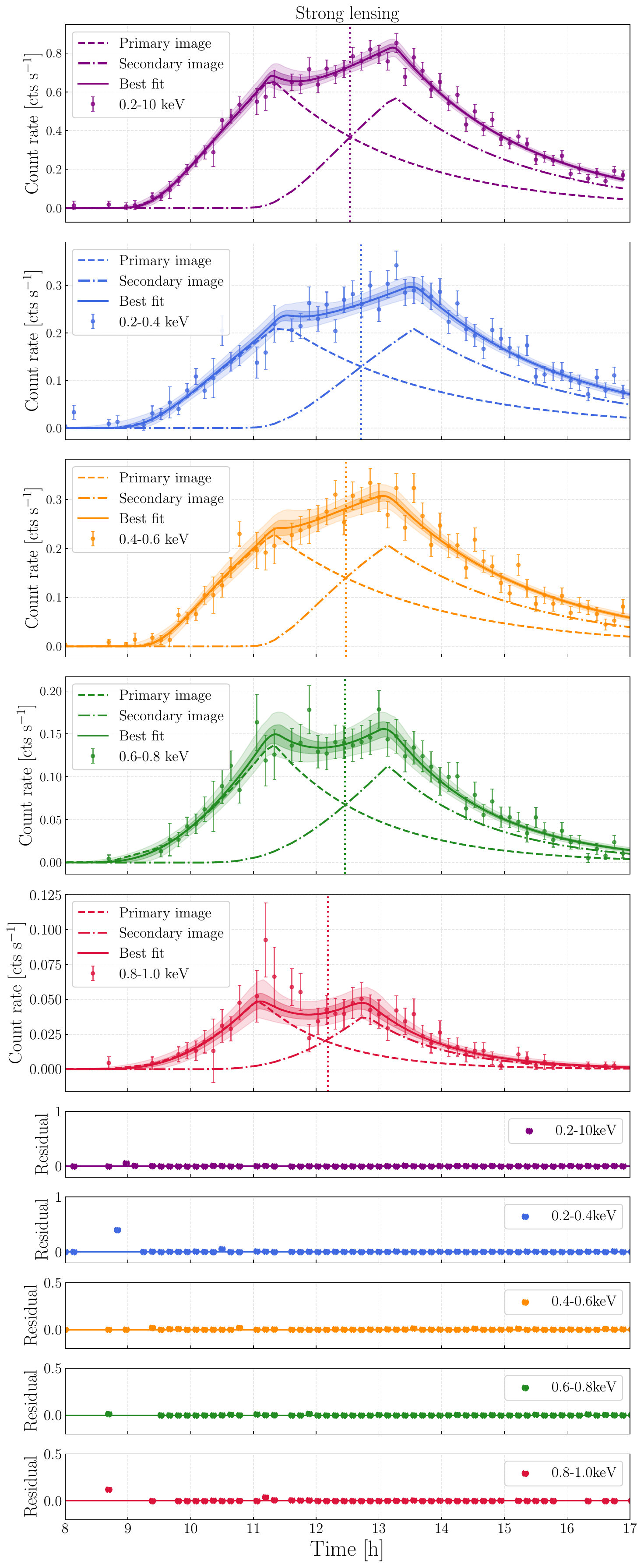}
\caption{The upper panel shows the best fitting results of eRO-QPE1 obs1 data in the energy ranges of $0.2$ - $10$ keV (purple), $0.2$ - $0.4$ keV (blue), $0.4$ - $0.6$ keV (orange), $0.6$ - $0.8$ keV (green), and $0.8$ - $1.0$ keV (red). In all five energy bands, the solid lines are derived from Equation. \eqref{eq20} combined with the posterior parameter distributions of the fitting results, the dashed lines from Equation. \eqref{eq18} combined with  the posterior parameter distributions of the fitting results, and the dot-dashed lines from Equation. \eqref{eq19} combined with the posterior parameter distributions of the fitting results. The lower panel shows the residual of each data point.}
\label{fig:total_fitted_image}
\end{figure}

\subsection{Light curve analysis and results}
\label{Light_curve_analysis}

The first observation of eRO-QPE1 was reported by \cite{2021Natur.592..704A}, who conducted a detailed analysis of the chaotic mixture of multiple overlapping eruptions with varying amplitudes \citep{2022A&A...662A..49A}. 
The corresponding light curve was decomposed into five energy bands and successfully fitted using a piecewise function \citep{2005ApJ...627..324N},
\begin{eqnarray}
f(t,A,\gamma_1,\gamma_2,t_{\rm peak})=\left\{
\begin{array}{rcl}
A \lambda e^{\gamma_1/(t_{\rm peak}-t_{\rm as}-t)} & & {t<t_{\rm peak}},\\
A e^{-(t-t_{\rm peak})/\gamma_2} & & {t \geq t_{\rm peak}}.
\end{array} \right.
\label{eq18}
\end{eqnarray}
Here $A$ is the amplitude of the eruptions, $t_{\rm as}=\sqrt{\gamma_1 \gamma_2}$, $\lambda=e^{t_{\lambda}}$ and $t_{\lambda}=\sqrt{\gamma_1/\gamma_2}$. 
Based on this formulation and in addition to the common features shared by QPEs, eRO-QPE1 exhibits a multi-component structure in its Obs1 data. We propose that gravitational lensing of the QPE signal could explain this complexity. 
In this interpretation, the profile given by Equation \eqref{eq18} corresponds to the primary image. 
The secondary image can then be described by applying a factor $\mu^{-1}$ and a time delay $\Delta t$ to the primary image. 
Thus, the profile for the secondary image is written as
\begin{eqnarray}
g(t,A,\mu,\gamma_1,\gamma_2,t_{\rm peak},\Delta t)=\mu^{-1} f(t-\Delta t,A,\gamma_1,\gamma_2,t_{\rm peak}).
\label{eq19}
\end{eqnarray}
It is straightforward to see that the secondary image arrives delayed by $\Delta t$ and is demagnified by a factor of $\mu^{-1}$. 
For the PM lens model, $\mu^{-1}$ is always less than 1, meaning the primary image is brighter than the secondary one \citep{2021ApJ...918L..34W}, a result consistent with the findings of \cite{2022A&A...662A..49A}. 
Consequently, in our fitting process, the parameter $\mu^{-1}$ is naturally constrained to the range $0 < \mu^{-1} < 1$. 
Therefore, based on Equations (\ref{eq18}) and (\ref{eq19}), the total flux can be expressed as

\begin{eqnarray}
\begin{aligned}
h(t,A,\mu,\gamma_1,\gamma_2,t_{\rm peak},\Delta t)=
f(t,A,\gamma_1,\gamma_2,t_{ \rm peak})\\
+g(t,A,\mu,\gamma_1,\gamma_2,t_{\rm peak},\Delta t).
\end{aligned}
\label{eq20}
\end{eqnarray}
We should also note that we may not observe two isolated images if the images lasting timescale is larger or comparable to the timescale of time delay. 

In the following, we adopt Equation \eqref{eq20} as our model and use it to fit the light curves in the following energy bands, using data provided by Arcodia: 0.2–0.4 keV, 0.4–0.6 keV, 0.6–0.8 keV, 0.8–1.0 keV, and 0.2–10 keV. 
Data in the 1.0–2.0 keV range are excluded due to their comparatively large error bars. 
In our fitting, we do not introduce an additional constant plateau component, instead, we focus solely on modeling the main pulse structure in the light curves. 
A flux threshold of approximately 0.001 cts s$^{-1}$ is applied to the data. 
For the time coordinate of the light curves, we applied a unit conversion and an offset to the original observation time. 
Specifically, across all energy bands, the time is transformed according to the relation: (Time / 3600) – 197850, which yields the time axis shown in the Figure \ref{fig:total_fitted_image}. 
Since we focused on fitting the two sub-eruptions within the main peak, which led us to apply a more constrained boundary to the time interval. 
Finally, the analyzed time interval for all light curves is restricted to 8–17 h.

We use the UltraNest \footnote{\url{https://johannesbuchner.github.io/UltraNest/}} package to fit the main pulse of Obs1 data \citep{2021JOSS....6.3001B,2016S&C....26..383B,2019PASP..131j8005B}.  
There are six parameters in our model, where $A$ represents the magnitude of the image, $\gamma_1$ and $\gamma_2$ represent the rise and decay timescale of the image, respectively. 
$t_{\rm peak}$ represents the peak time of one single image. 
The peak time of the secondary image is retarded $\Delta t$ times later and its flux is de-magnified by a factor of $\mu^{-1}$. 
As summarized in the second column of Table \ref{tab:best_fit}, the adopted parameter ranges are informed by the following considerations. 
For parameter $A$, the upper bound is set above the peak flux, while the lower bound is taken as approximately half of the peak flux. 
The parameter $t_{\text{peak}}$ is constrained to a range centered on the time of peak flux and spanning approximately $\pm 1$ hour. 
Both $\gamma_1$ and $\gamma_2$ are constrained within 0.01–15 h. 
Given the slightly fainter second sub-eruption and the lensing hypothesis, the magnification $\mu^{-1}$ is limited to the interval $[0.5, 0.99]$. 
Based on the results shown in Figure 6 of  \cite{2022A&A...662A..49A}, the separation between the two peak times is around 2 h; accordingly, the parameter $\Delta t$ is set between 1 and 3 h.

The best-fit result is given in the third column of Table \ref{tab:best_fit}. 
The fitting result is shown in Figure \ref{fig:total_fitted_image}. 
The upper panel shows the fitting results in the energy ranges of $0.2$ - $10$ keV (purple), $0.2$ - $0.4$ keV (blue), $0.4$ - $0.6$ keV (orange), $0.6$ - $0.8$ keV (green), and $0.8$ - $1.0$ keV (red). 
The relative dark shade region represents their uncertainty, which is the same in all five energy ranges. 

Combining the fitting result in Table \ref{tab:best_fit}, the two sub-eruptions and the sum of them are plotted in Figure \ref{fig:total_fitted_image} with three types of lines.  
The dashed line and dash-dotted line in each energy band represent the primary image (e.g., first sub-eruption) and the secondary image (e.g., second sub-eruption), respectively. 
While the solid line in each energy band represents the sum of the two images. 
To quantify the discrepancies between the model and the observations, the residuals are calculated using  $\chi^2 _i = (x_{t_{\rm i}} - m_{t_{\rm i}})^2/m_{t_{\rm i}}$, where $x_{t_{\rm i}}$ is the observed data point at time $t_{\rm i}$ and ${m_{t_{\rm i}}}$ is the corresponding model value \citep{pearson1900x}. 
The lower panel of Figure \ref{fig:total_fitted_image} shows the residuals of the data points in the five energy ranges. 
To facilitate the subsequent spectral analysis and in consideration of the characteristic fast rise and slow decay of the light curve, we define a time point—referred to as the cross-time—based on the fitting results. 
This cross-time is determined by the intersection between the decay of the first sub-eruption and the rise of the second sub-eruption. 
Before the cross-time, the first sub-eruption dominates, whereas after the cross-time, the second sub-eruption becomes dominant. 
The cross-times for the five energy bands are marked in Figure \ref{fig:total_fitted_image} with vertical dotted lines.

\subsection{Spectral analysis and result}

In order to investigate the X-ray emission properties, we analyzed the data from Obs1 (with a total exposure of $\sim$94 ks). The XMM-Newton EPIC pn camera was operated in small-window mode for this observation. We processed the data using the XMM-Newton Science Analysis Software (SAS), running the standard pipeline tasks epproc to produce calibrated event lists for the EPIC pn and MOS cameras. 
Based on the cross-time determined in Section \ref{Light_curve_analysis}, we defined the following time intervals for the spectral analysis of each energy band. 
As the cross-time has already separated the two sub-eruptions, we did not impose strict constraints on the boundary values of the time range during the spectral analysis. 
For the 0.2–10 keV band, the first and second sub-eruption span approximately 5–12.5 h and 12.5–18 h, respectively. 
In the 0.2–0.4 keV band, the intervals are approximately 5–12.7 h for the first sub-eruption and 12.7–18 h for the second sub-eruption. 
For the 0.4–0.6 keV band, they are approximately 5–12.5 h and 12.5–18 h. 
In the 0.6–0.8 keV band, the corresponding ranges are approximately 5–12.5 h and 12.5–18 h. 
Finally, for the 0.8–1.0 keV band, the first sub-eruption covers approximately 5–12.3 h and the second sub-eruption covers approximately 12.3–18 h. 
These intervals are then used for the subsequent spectral extraction and analysis. 
For the spectral analysis of the two sub-eruption time intervals, the background is extracted from a nearby source-free region, and the response matrices and ancillary response files are generated using the XMMSAS tasks rmfgen and arfgen. 
The spectral channels are grouped to achieve a signal-to-noise ratio (S/N) of 3 in each energy bin. 

The spectra are then fitted with a power-law model using XSPEC (version 12.12.1) in the energy ranges of $0.2$–$0.4$ keV, $0.4$–$0.6$ keV, $0.6$–$0.8$ keV, and $0.8$–$1.0$ keV, while the $0.2$–$10$ keV band is fitted with the $tbabs$ and $diskbb$ model. 
The best-fit spectrum is presented in Figure \ref{spec-all} and \ref{spec-all-b}. 
Figure \ref{fig:photon_index} illustrates the evolution of the photon index across the four energy bands for the two sub-eruptions. 
In the 0.2-10 keV energy range, the fitting parameters for the first sub-eruption, using the $tbabs$ and $diskbb$ models, are as follows: the hydrogen column density is approximately ${1.2^{+0.01}_{-0.01} \times 10^{21}} $ cm$^{-2}$, and the disk blackbody temperature is approximately $0.12^{+0.004}_{-0.004}$ keV. 
For the second sub-eruption, the fitting parameters are: the hydrogen column density is approximately $1.0^{+0.08}_{-0.08} \times 10^{21}$ cm$^{-2}$, and the disk blackbody temperature is approximately $0.10^{+0.002}_{-0.002}$ keV. 
We note that the photon index tends to increase with energy. 
Furthermore, a comparison of the photon index, the hydrogen column density, and the disk blackbody temperature between the two sub-eruption reveals no significant spectral differences, suggesting that the emission mechanisms are likely similar in the two sub-eruptions. 
This supports the interpretation that the two sub-eruptions may originate from the same underlying physical process.

\subsection{Lensing parameters}
According to the properties of gravitational lensing, the posterior distributions of time delays and magnification ratios should be consistent across different energy bands. 
In Figure \ref{fig:energyVStimedelay}, we present the time delays and magnification for five energy bands. 
The fitting results show that the time delay is approximately $1.9$ hours, and the magnification ratio is about $1.1$. 
Based on the value of $\mu$, we can further estimate the flux of obs1 in the absence of lensing. 
For the PM model, substituting $\mu \approx 1/0.9$ into Equation \eqref{eq9} yields a value of $y \approx 0.1$. 
Inserting this $y$ into Equation \eqref{eq8} gives $\mu_+ \approx 5.5$ and $\mu_{-} \approx 4.5$. 
This indicates that under the PM model, the lensing effect amplifies the eruption of obs1 by approximately a factor of 5. 
For the SIS model, Equation \eqref{eq16} gives $\mu = \mu_+/\mu_-$. 
Substituting $\mu \approx 1/0.9$ yields $y \sim 0.1$. 
With $y \approx 0.1$, Equation \eqref{eq16} gives $\mu_+ \sim 11$ and $\mu_- \sim 9$. 
Similar to other QPEs \citep{2025A&A...703A.263H}, eRO-QPE1 shows a sequence of eruptions that typically fade, brighten, and then gradually fade away \citep{2024ApJ...965...12C}. 
This pattern suggests that obs1 might be an early eruption in the eRO-QPE1 sequence. 
At this point, the lensing model still remains some consistency with the observations.

Based on the posterior distributions of the time delay and magnification from the fitting result (see Table \ref{tab:best_fit}). 
One can estimate the lens mass across the five energy bands. 
For the PM model, $M_{\rm Lz} \approx 7.53_{-4.35}^{+10.86} \times 10^{9} M_{\odot}$ in the $0.2$ - $0.4$ keV band, $M_{\rm Lz} \approx 3.16^{+8.25}_{-1.65} \times 10^{9} M_{\odot}$ in $0.4$ – $0.6$ keV band, $M_{\rm Lz} \approx 1.60^{+4.25}_{-0.86} \times 10^{9} M_{\odot}$ in $0.6$ – $0.8$ keV band, $M_{\rm Lz} \approx 1.25^{+1.61}_{-0.37} \times 10^{9} M_{\odot}$ in $0.8$ – $1.0$ keV band, and $M_{\rm Lz} \approx 2.38^{+2.02}_{-0.81} \times 10^{9} M_{\odot}$ in $0.2$ – $10$ keV band. 
For the SIS model, $M_{\rm Lz} \approx 7.53_{-4.35}^{+10.86} \times 10^{9} M_{\odot}$ in the $0.2$ - $0.4$ keV band, $M_{\rm Lz} \approx 3.16^{+8.25}_{-1.65} \times 10^{9} M_{\odot}$ in $0.4$ – $0.6$ keV band, $M_{\rm Lz} \approx 1.61^{+2.65}_{-0.74} \times 10^{9} M_{\odot}$ in $0.6$ – $0.8$ keV band, $M_{\rm Lz} \approx 1.26^{+1.60}_{-0.37} \times 10^{9} M_{\odot}$ in $0.8$ – $1.0$ keV band, and $M_{\rm Lz} \approx 2.39^{+2.02}_{-0.81} \times 10^{9} M_{\odot}$ in $0.2$ – $10$ keV band. 
The estimated values across the different energy bands remain consistent within their substantial statistical uncertainties. 
The observed scatter likely stems from the propagation of fitting errors in the derived time delays and magnification ratios, to which the mass estimates are particularly sensitive. Thus, the overall dispersion appears compatible with the wavelength-independent lensing assumption. 

Due to the fact that the complex structure of the light curve in obs1 does not appear in every other eruptions, we investigate here the Einstein timescale of the lens. 
The expression for the Einstein timescale is: $t_0 = r_{\rm E} / v$, where $v$ is the tangential velocity between the source and the lens. 
Therefore, the Einstein timescale is jointly determined by the mass of the lens, the relative distances among the observer, the lens, and the source, as well as the velocity $v$. 
The known quantity is that the distance to the source is approximately $200$ Mpc ($z \approx 0.05$) and the central mass of the host galaxy is $M_{\rm SMBH} \sim 10^6M_{\odot}$. 
For the PM lens model, we consider two approximate scenarios. 
Regardless of whether $D_{\rm L} \sim 0.5 D_{\rm S}$ or $D_{\rm S} \sim D_{\rm L}$, even with extreme parameters, the results cannot be reconciled with the observations. 
For example, under the approximation of $D_{\rm L} \approx D_{\rm S}$, we have $r_{\rm E} \approx \sqrt{2R_{\rm S} D_{\rm LS}}$. 
Assuming a lens mass of $10^8 M_{\odot}$, $D_{\rm LS} \approx 10^7 R_{\rm S}(M_{\rm L})$, and a relative velocity between the source and lens of $0.1c$, under these extreme parameters, $t_0 \sim 1$ year. 
Under the approximation of $D_{\rm L} = 0.5 D_{\rm S}$, while keeping all other parameters the same, $t_0 \sim 1000$ years under these extreme parameter conditions. 
For the SIS lens model, the Einstein timescale is determined by the relative distances, the velocity dispersion $\sigma_v$, and the relative tangential velocity $v$. 
Using the approximations $D_{\rm L} \approx D_{\rm S}$ and $D_{\rm LS} \approx 10^6 R_{\rm S}(M_{\rm SMBH})$, with $M_{\rm SMBH} \sim 10^6 M_{\odot}$, $v = \sqrt{G M_{\rm SMBH} / D_{\rm LS}}$, and $\sigma_v \sim 50 \ \rm{km \ s^{-1}}$, the Einstein timescale under these parameter settings is approximately $7$ hours. 
If $D_{\rm LS}$ is set within the range of $10^6 R_{\rm S}(M_{\rm SMBH})$ to $10^7 R_{\rm S}(M_{\rm SMBH})$, the Einstein timescale ranges from several hours to several days. 
While this conclusion is not inconsistent with the observations, the approximation $D_{\rm L} \approx D_{\rm S}$ implies that the source might need to be located at the outskirts of a dwarf galaxy (with $z_L \sim 0.05$). 
The concentration of the host galaxy's mass within the range of $10^{6-7}R_{\rm S}(M_{\rm SMBH})$ poses a challenge to the SIS model. 
Further constraining the lens system would require calculations incorporating the projected surface mass density profile of the dwarf galaxy along the line of sight \citep{1992ARA&A..30..311B}, which we did not pursue further at this stage.

\begin{figure*}
\centering
\setlength{\tabcolsep}{0.8pt}
\begin{tabular}{@{}c c@{}}
\includegraphics[width=0.45\textwidth,angle=0]{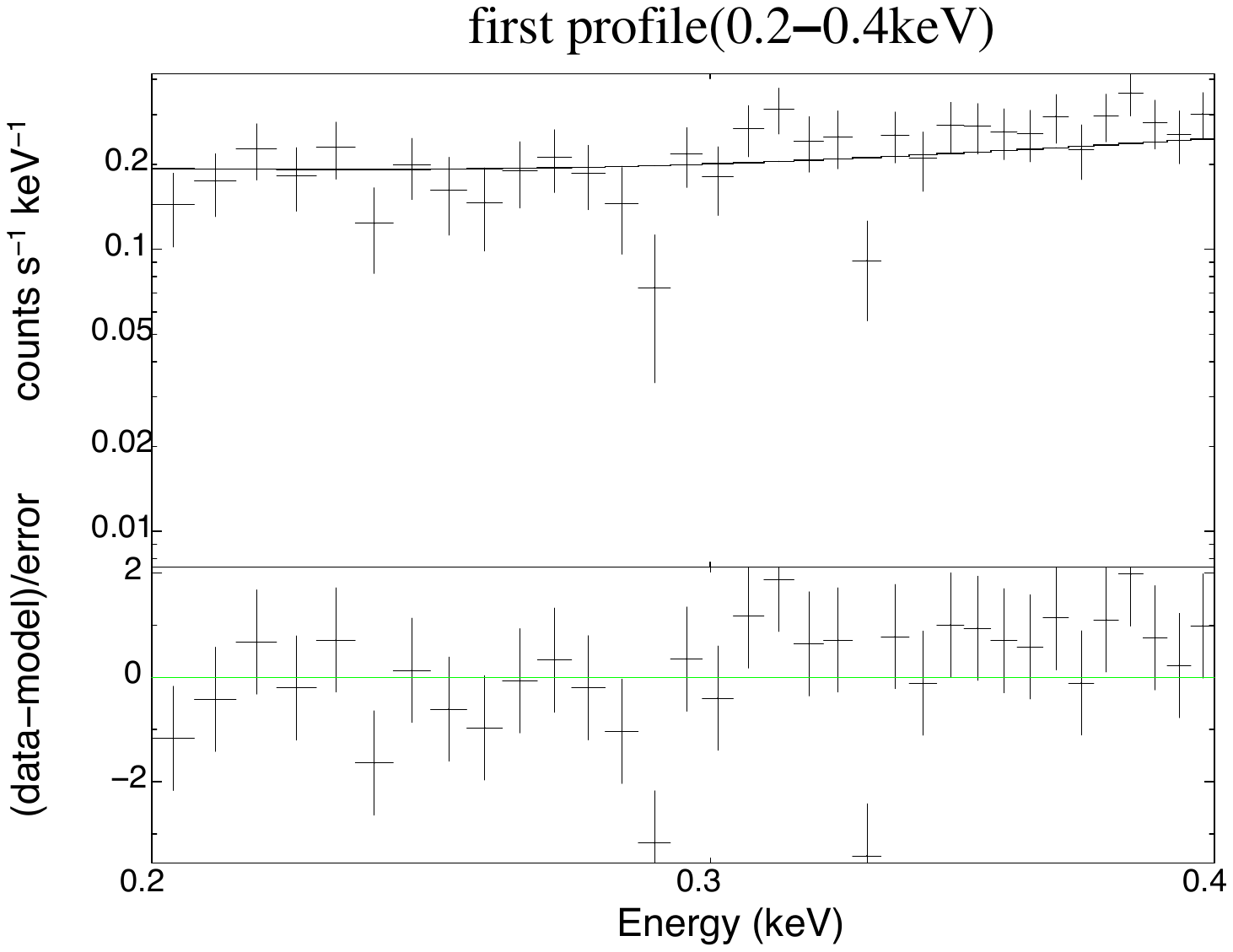} &
\includegraphics[width=0.45\textwidth,angle=0]{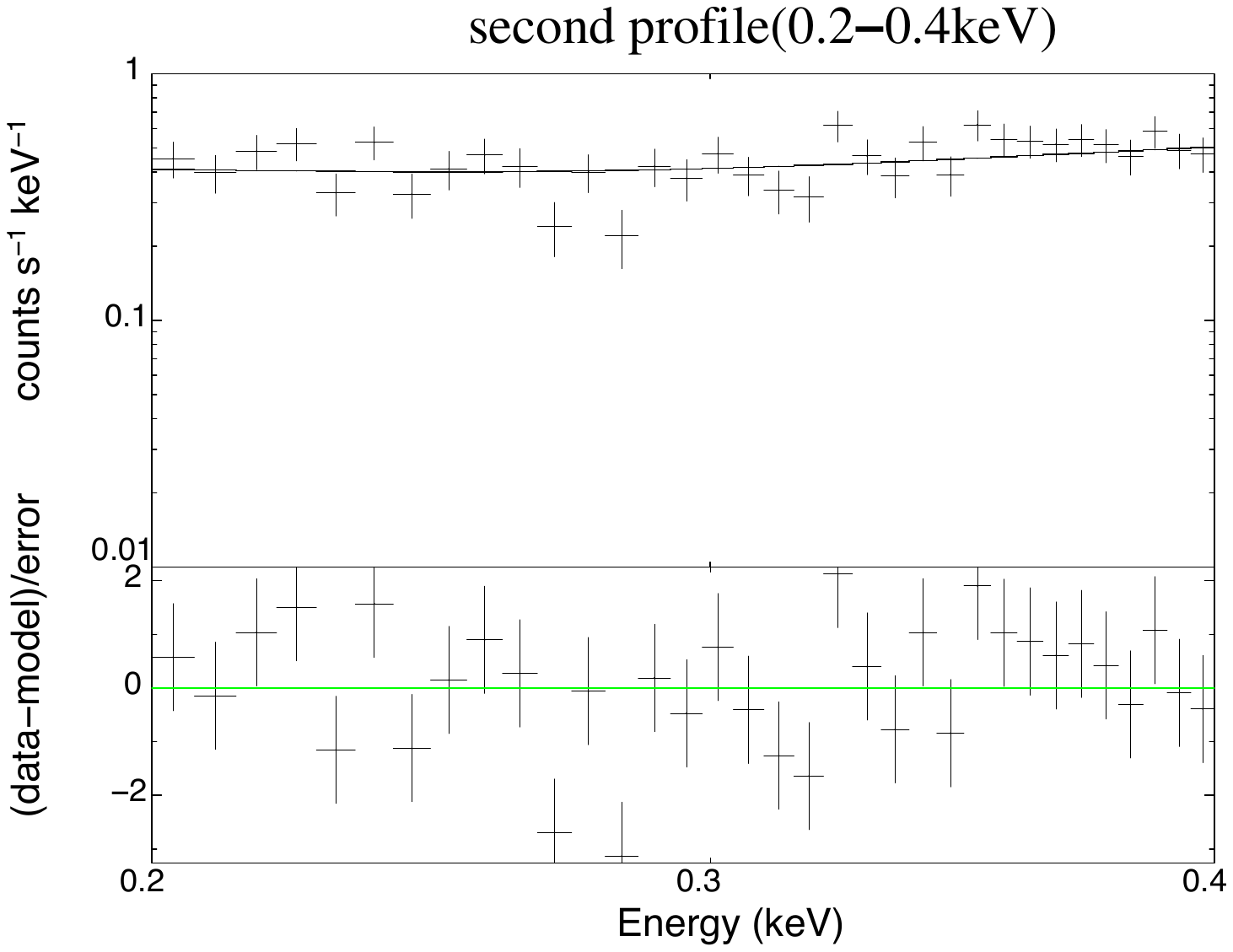} \\
& \\[-8ex]
\includegraphics[width=0.45\textwidth,angle=0]{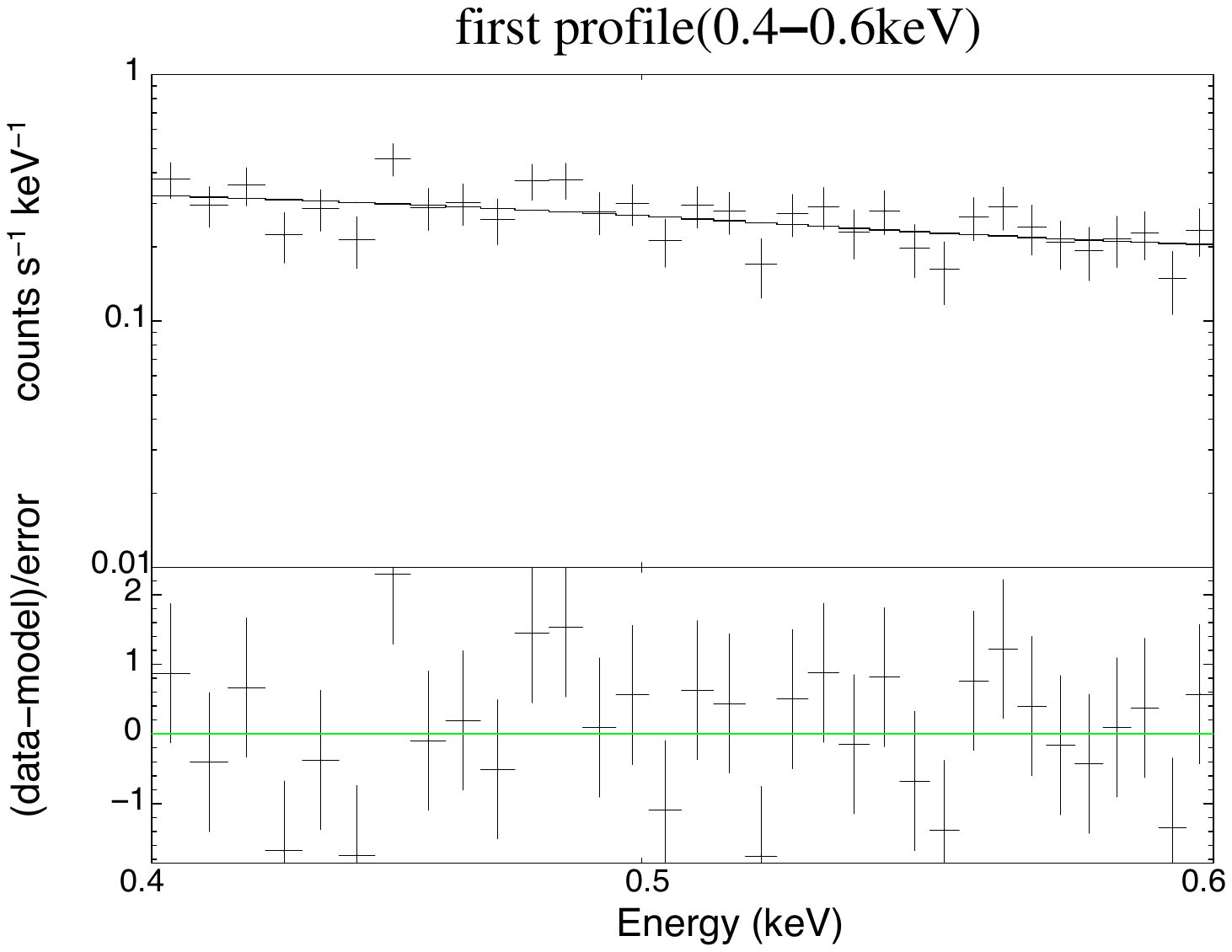} &
\includegraphics[width=0.45\textwidth,angle=0]{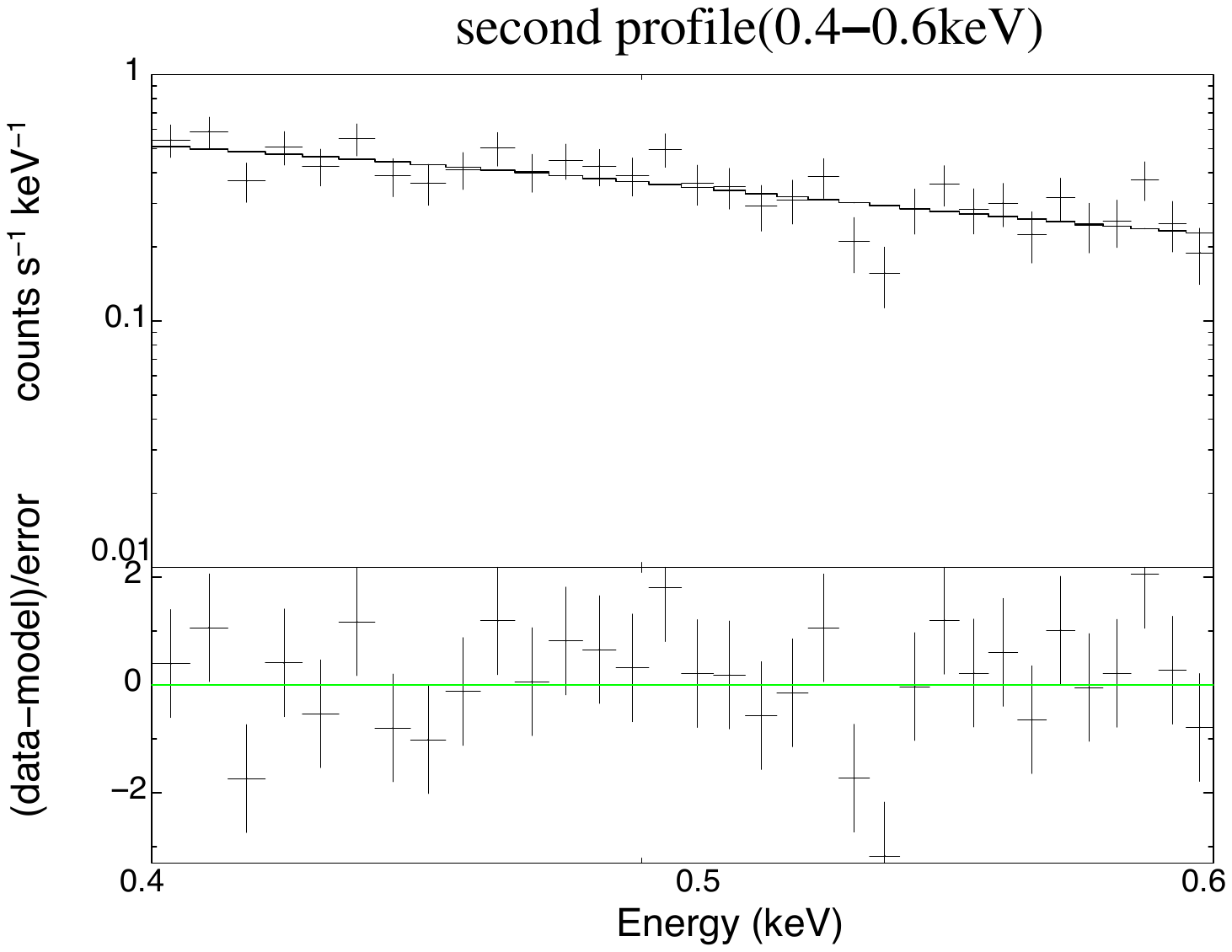} \\
& \\[-8ex]
\includegraphics[width=0.45\textwidth,angle=0]{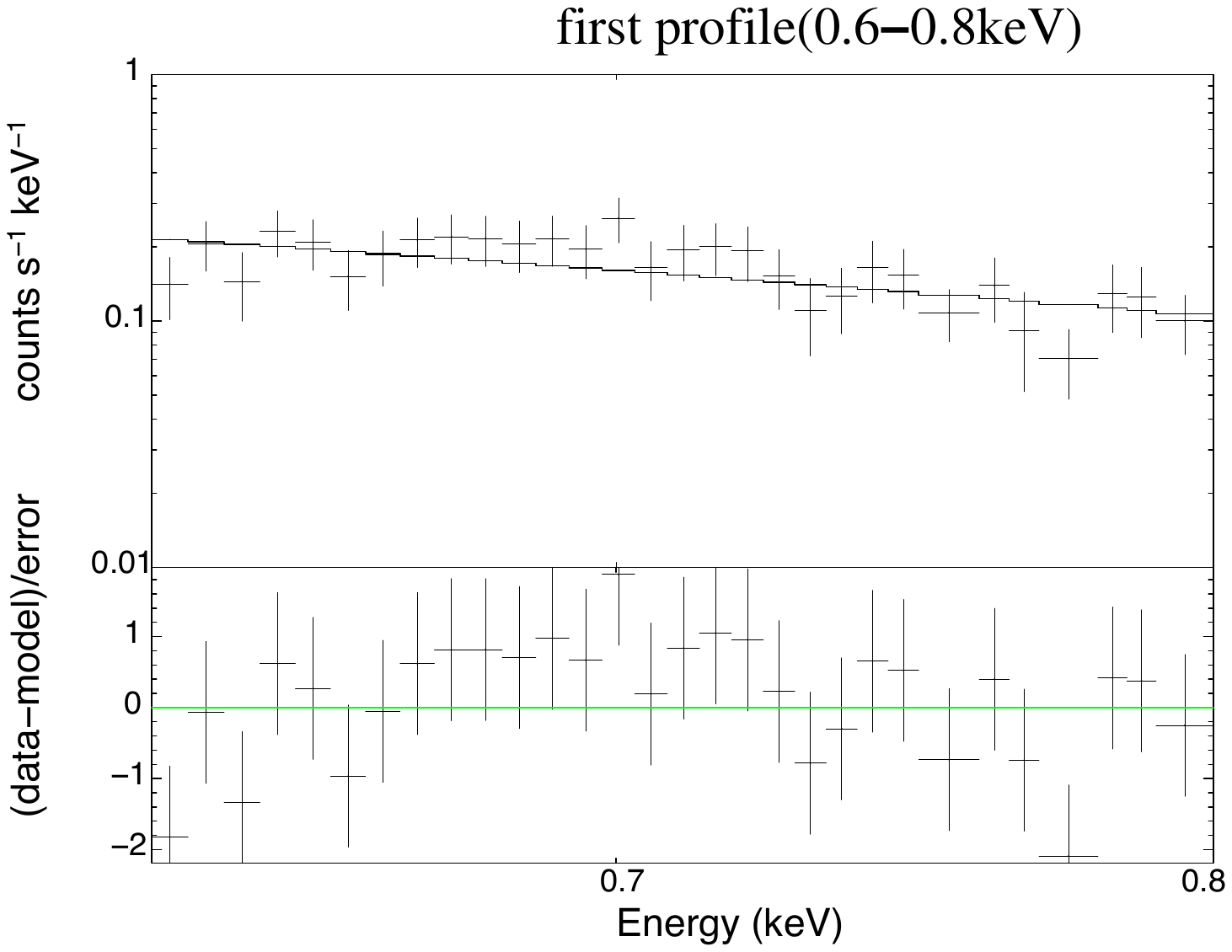} &
\includegraphics[width=0.45\textwidth,angle=0]{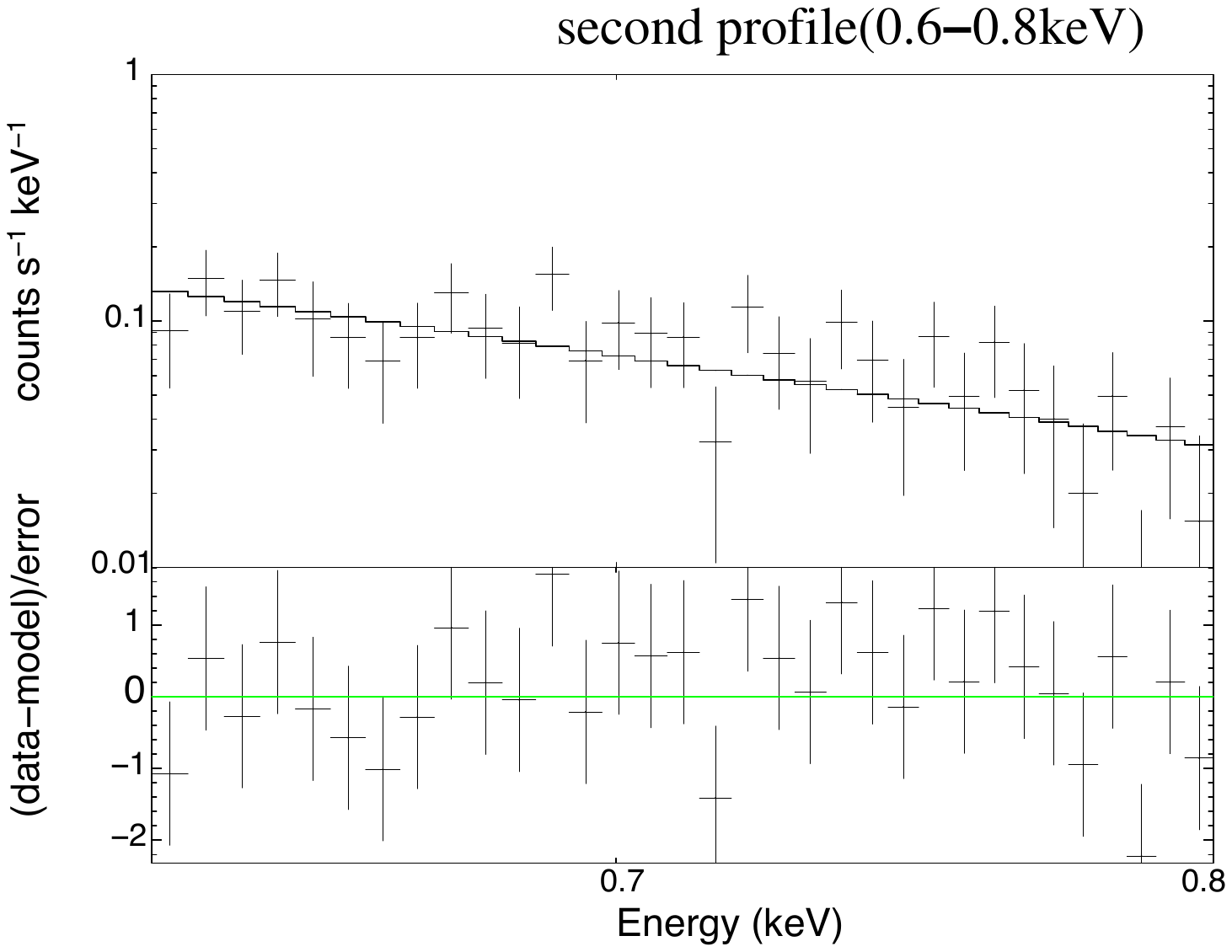} \\
& \\[-8ex]
\includegraphics[width=0.45\textwidth,angle=0]{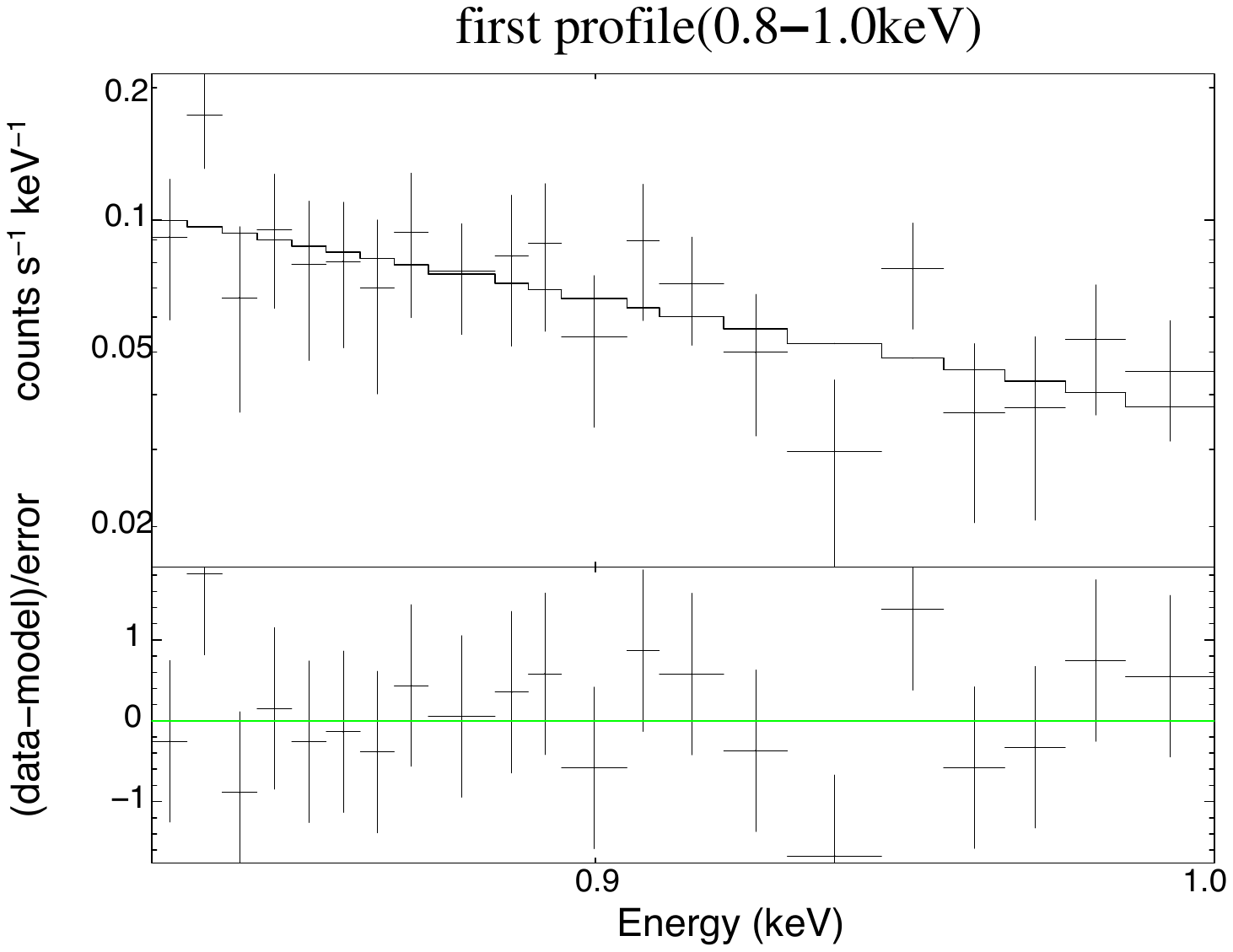} &
\includegraphics[width=0.45\textwidth,angle=0]{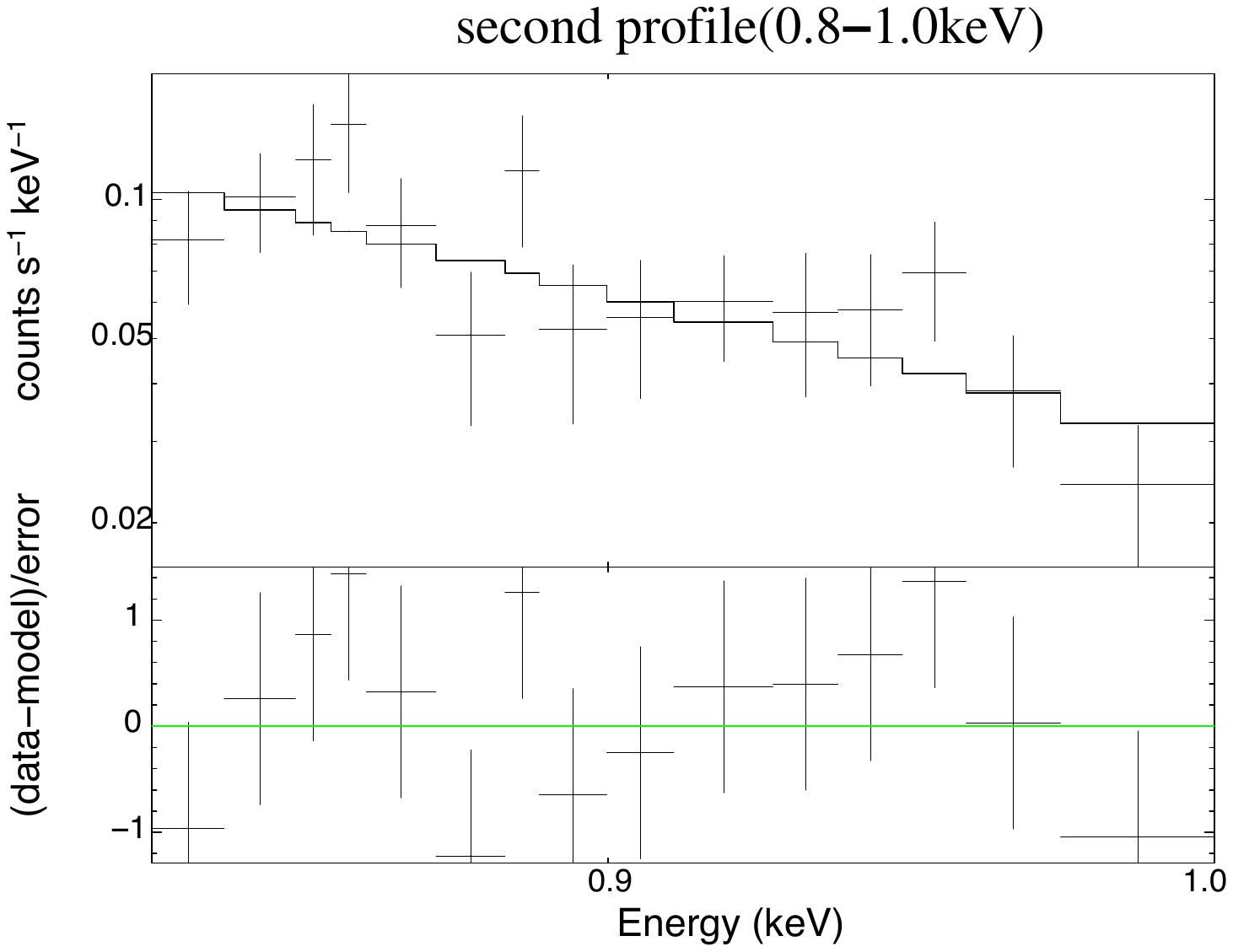} \\
\end{tabular}
\caption{The phase-resolved spectra of obs1 in the energy bands of 0.2-0.4 keV, 0.4-0.6 keV, 0.6-0.8 keV and 0.8-1.0keV are presented. 
The left and right columns show the first and second sub-eruptions (or profile), respectively. 
These four energy bands are fitted with a power-law model. (Continued on Figure \ref{spec-all-b})}
\label{spec-all}
\end{figure*}

\begin{figure*}
\centering
\setlength{\tabcolsep}{0.8pt}
\begin{tabular}{@{}c c@{}}
\includegraphics[width=0.45\textwidth,angle=0]{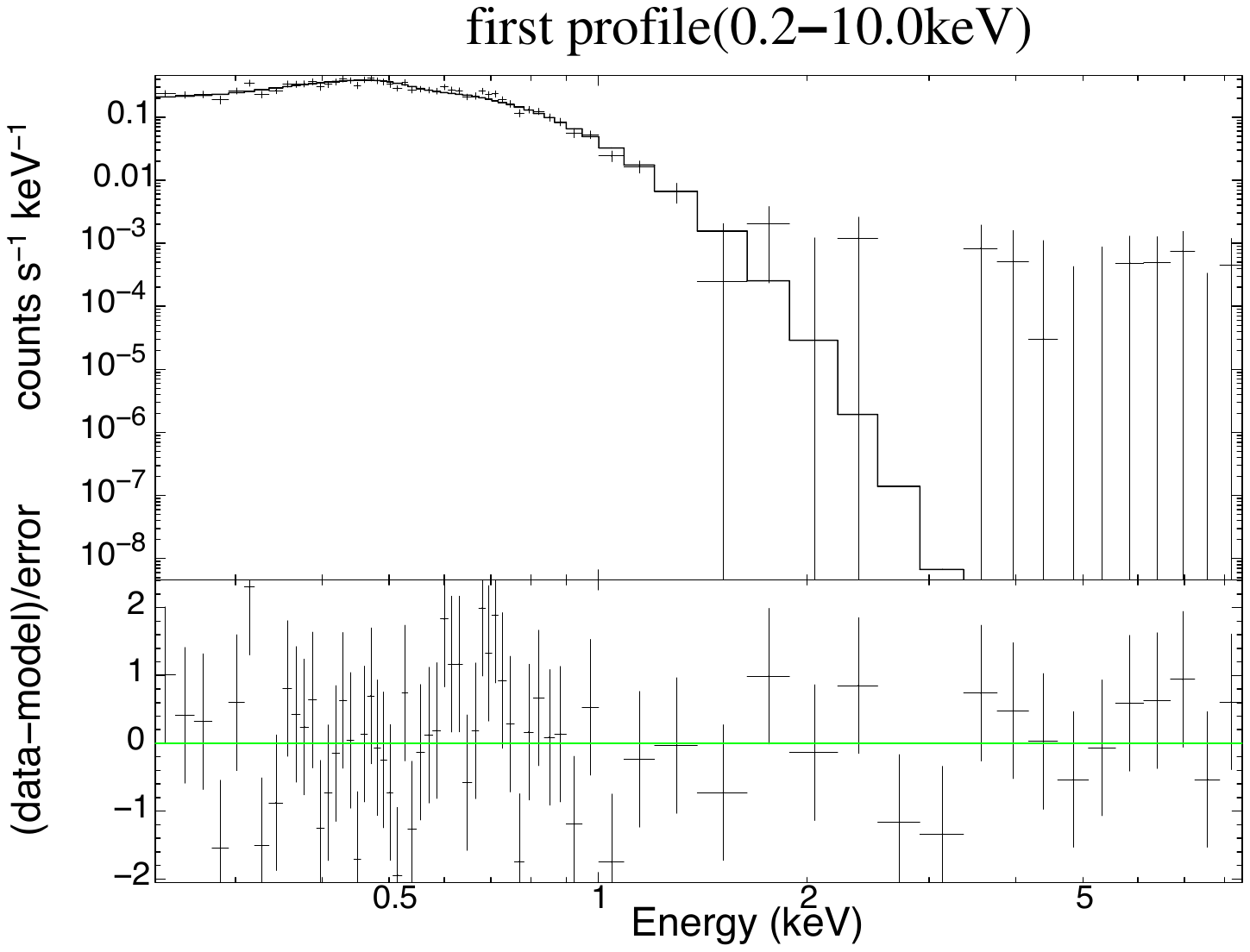} &
\includegraphics[width=0.45\textwidth,angle=0]{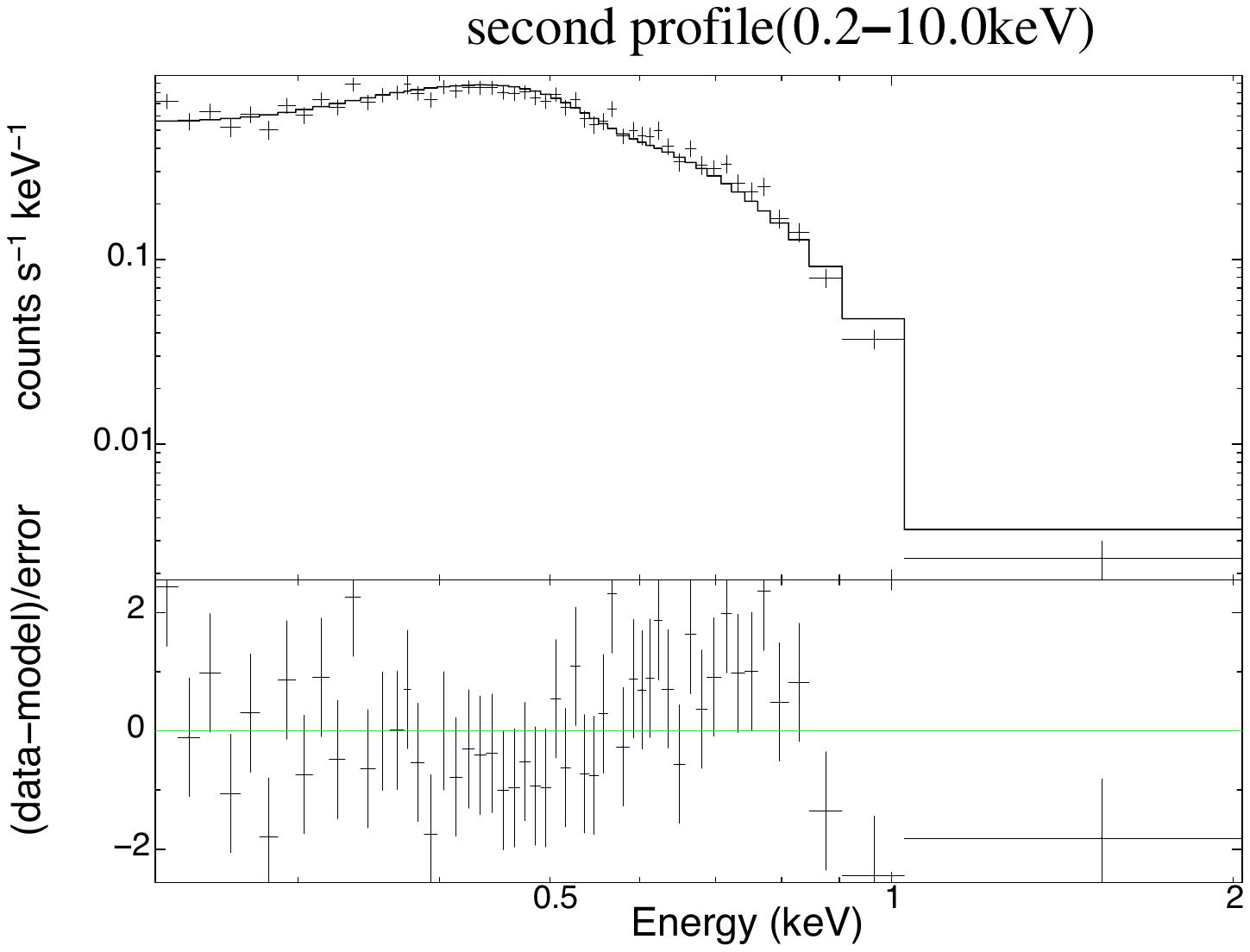} \\
\end{tabular}
\caption{(Continued from Figure \ref{spec-all}) The phase-resolved spectra of obs1 in the energy band of 0.2-10 keV are presented. 
The left and right columns show the first and second sub-eruptions (or profile), respectively. 
The 0.2-10 keV band is fitted with the $tbabs$ and $diskbb$ model.}
\label{spec-all-b}
\end{figure*}

\begin{figure}  
\centering
\includegraphics[width=8.5cm]{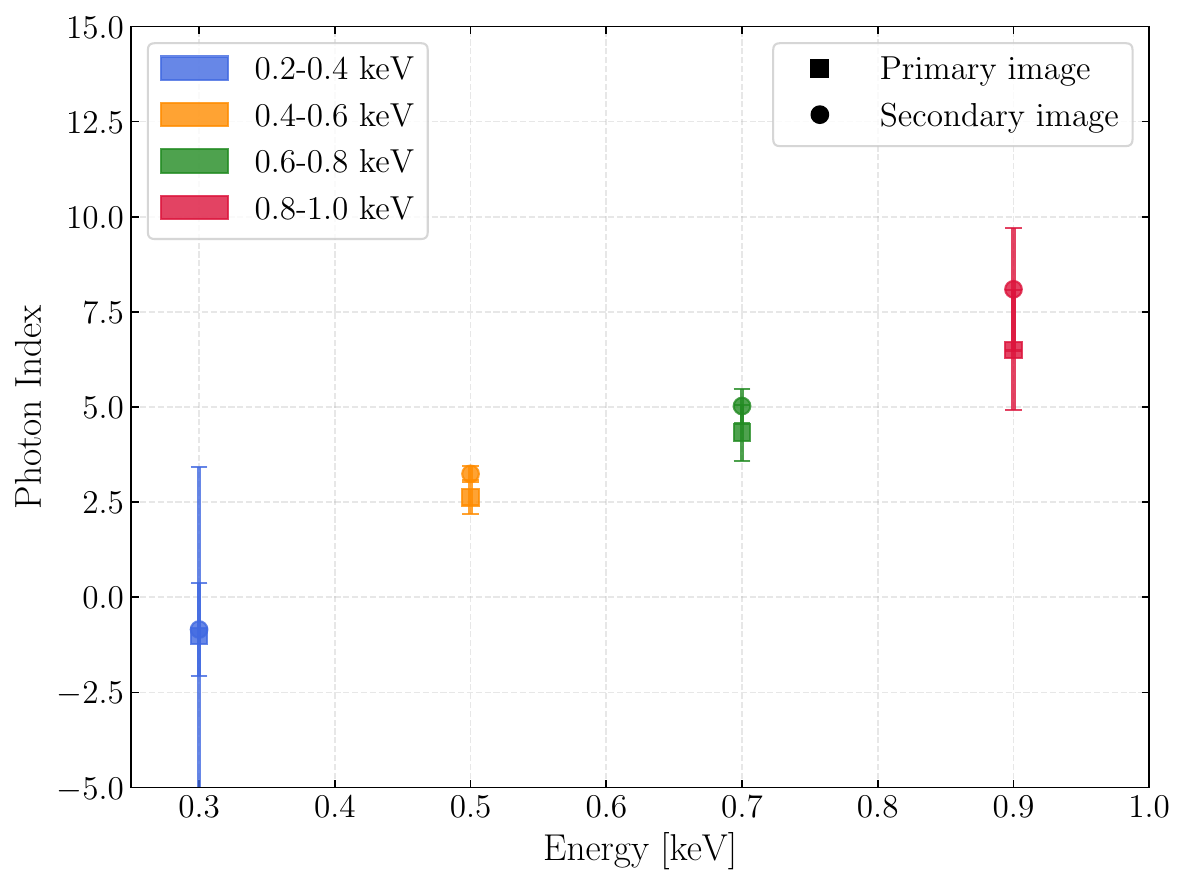}
\caption{Temporal evolution of the spectral power-law index compared for the two sub-eruptions. The photon index of the first sub-eruptions are represented by squares, while those of the second sub-eruptions are denoted by circles. Different energy bands are color-coded as follows: 0.2–0.4 keV (blue), 0.4–0.6 keV (orange), 0.6–0.8 keV (green), and 0.8–1.0 keV (red).}
\label{fig:photon_index}
\end{figure}

\begin{figure}  
\centering
\includegraphics[width=8cm]{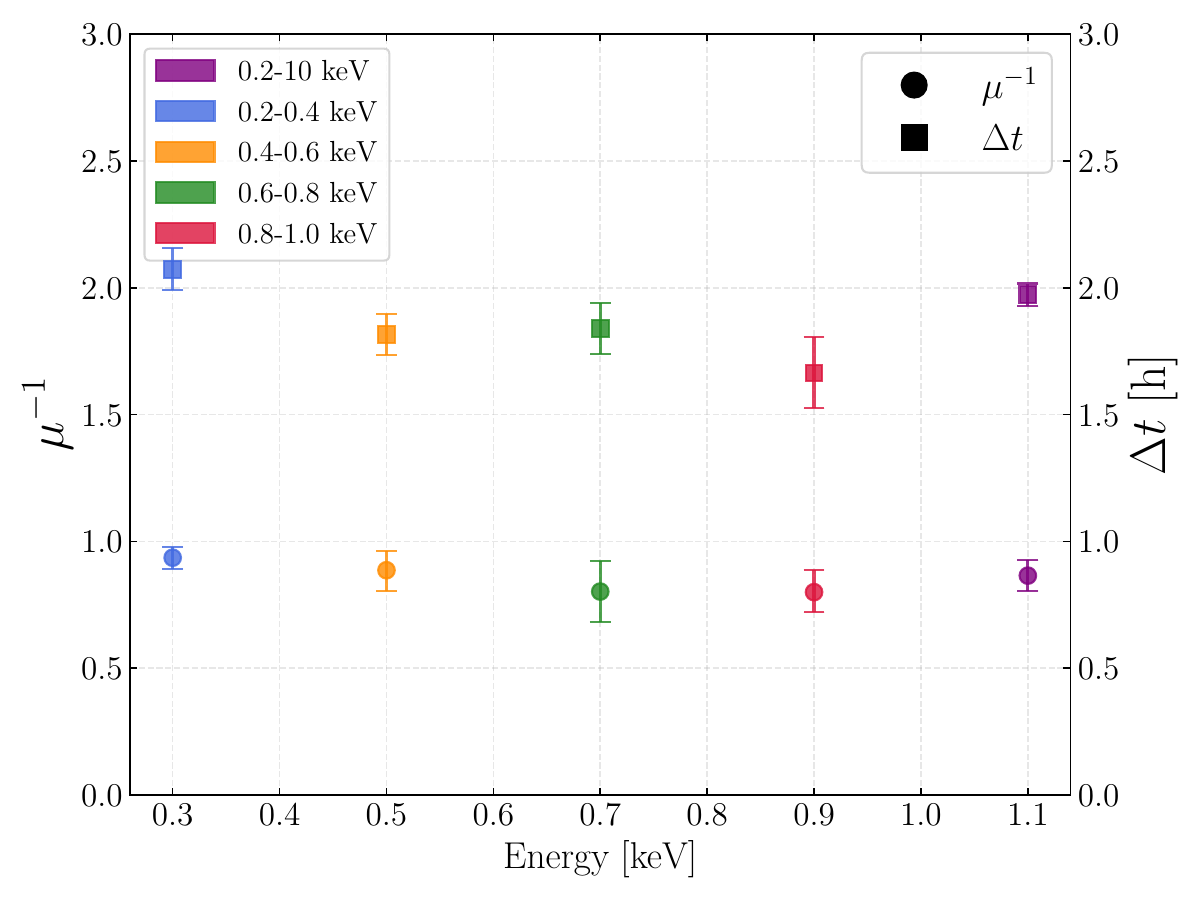}
\caption{The median posterior estimates of time delays (squares) and magnifications (circles) across the five energy bands. Different energy bands are color-coded as follows: 0.2–10 keV (purple), 0.2–0.4 keV (blue), 0.4–0.6 keV (orange), 0.6–0.8 keV (green), and 0.8–1.0 keV (red).}
\label{fig:energyVStimedelay}
\end{figure}

\renewcommand{\arraystretch}{1.4}
\begin{table}
\setlength\tabcolsep{8pt} 
\centering
\caption{Parameter space and best-fit values for the energy ranges of $0.2$–$0.4$ keV, $0.4$–$0.6$ keV, $0.6$–$0.8$ keV, $0.8$–$1.0$ keV, and $0.2$–$10$ keV.}
\label{tab:best_fit}
\begin{tabular}{cccc}
\hline
\hline
Parameter&Parameter range&Best-Fit\\
\hline
\hline
$A_{0.2-0.4\rm{keV}} \ [\rm{cts\ s^{-1}}]$&$[0.1-0.3]$&$0.22_{-0.01}^{+0.01}\ $\\
\hline
$t_{p_{0.2-0.4\rm{keV}}}\ [\rm{h}]$&$[10, 13]$&$11.45_{-0.07}^{+0.09}$\\
\hline
$\gamma_{1_{0.2-0.4\rm{keV}}}\ [\rm{h}]$&$[0.01, 15]$&$4.21_{-0.89}^{+1.5}$\\
\hline
$\gamma_{2_{0.2-0.4\rm{keV}}}\ [\rm{h}]$&$[0.01, 15]$&$2.38_{-0.13}^{+0.13}$\\
\hline
$\mu^{-1}_{0.2-0.4\rm{keV}}$&$[0.5,0.99]$&$0.95_{-0.06}^{+0.03}$\\ 
\hline
$\Delta t _{0.2-0.4\rm{keV}}\ [\rm{h}]$&$[1,3]$&$2.11_{-0.09}^{+0.08}$\\ 
\hline
\hline
$A_{0.4-0.6\rm{keV}}\ [\rm{cts\ s^{-1}}]$&$[0.1,0.3]$&$0.23_{-0.01}^{+0.01}$\\
\hline
$t_{p_{0.4-0.6\rm{keV}}} \ [\rm{h}]$&$[10, 13]$&$11.32_{-0.08}^{+0.10}$\\
\hline
$\gamma_{1_{0.4-0.6\rm{keV}}} \ [\rm{h}]$&$[0.01, 15]$&$2.71_{-0.55}^{+0.77}$\\
\hline
$\gamma_{2_{0.4-0.6\rm{keV}}} \ [\rm{h}]$&$[0.01, 15]$&$2.32_{-0.11}^{+0.11}$\\
\hline
$\mu^{-1}_{0.4-0.6\rm{keV}}$&$[0.5,0.99]$&$0.90_{-0.09}^{+0.07}$\\ 
\hline
$\Delta t _{0.4-0.6\rm{keV}} \ [\rm{h}]$&$[1,3]$&$1.82_{-0.09}^{+0.08}$\\ 
\hline
\hline
$A_{0.6-0.8\rm{keV}} \ [\rm{cts \ s^{-1}}]$&$[0.05-0.2]$&$0.14_{-0.01}^{+0.01}$\\
\hline
$t_{p_{0.6-0.8\rm{keV}}} \ [\rm{h}]$&$[10, 12]$&$11.30_{-0.10}^{+0.10}$\\
\hline
$\gamma_{1_{0.6-0.8\rm{keV}}} \ [\rm{h}]$&$[0.01, 15]$&$9.17_{-3.06}^{+3.72}$\\
\hline
$\gamma_{2_{0.6-0.8\rm{keV}}} \ [\rm{h}]$&$[0.01, 15]$&$1.60_{-0.07}^{+0.08}$\\
\hline
$\mu^{-1}_{0.6-0.8\rm{keV}}$&$[0.5,0.99]$&$0.81_{-0.12}^{+0.11}$\\ 
\hline
$\Delta t_{0.6-0.8\rm{keV}} \ [\rm{h}]$&$[1,3]$&$1.85_{-0.11}^{+0.09}$\\ 
\hline
\hline
$A_{0.8-1.0\rm{keV}} \ [\rm{cts\ s^{-1}}]$&$[0.02-0.1]$&$0.05_{-0.01}^{+0.01}$\\
\hline
$t_{p_{0.8-1.0\rm{keV}}} \ [\rm{h}]$&$[10, 12]$&$11.08_{-0.10}^{+0.08}$\\
\hline
$\gamma_{1_{0.8-1.0\rm{keV}}} \ [\rm{h}]$&$[0.01, 15]$&$11.35_{-3.66}^{+2.63}$\\
\hline
$\gamma_{2_{0.8-1.0\rm{keV}}} \ [\rm{h}]$&$[0.01, 15]$&$1.18_{-0.13}^{+0.14}$\\
\hline
$\mu^{-1}_{0.8-1.0\rm{keV}}$&$[0.5,0.99]$&$0.78_{-0.06}^{+0.11}$\\ 
\hline
$\Delta t _{0.8-1.0\rm{keV}} \ [\rm{h}]$&$[1,3]$&$1.70_{-0.12}^{+0.12}$\\ 
\hline
\hline
$A_{0.2-10\rm{keV}} \ [\rm{cts \ s^{-1}}]$&$[0.3-0.9]$&$0.67_{-0.03}^{+0.03} \ $\\
\hline
$t_{p_{0.2-10\rm{keV}}} \ [\rm{h}]$&$[10, 13] $&$11.28_{-0.05}^{+0.05}$\\
\hline
$\gamma_{1_{0.2-10\rm{keV}}} \ [\rm h]$&$[0.01, 15]$&$3.68_{-0.49}^{+0.59}$\\
\hline
$\gamma_{2_{0.2-10\rm{keV}}} \ [\rm h]$&$[0.01, 15]$&$2.10_{-0.07}^{+0.07}$\\
\hline
$\mu^{-1}_{0.2-10\rm{keV}}$&$[0.5,0.99]$&$0.86_{-0.06}^{+0.06}$\\ 
\hline
$\Delta t _{0.2-10\rm{keV}} \ [\rm h]$&$[1,3]$&$1.97_{-0.05}^{+0.04}$\\ 
\hline
\hline
\end{tabular}
\end{table}

\section{Optical depth}
\label{sec:optical_depth}
In this section, we investigate the probability of lensed QPE.
The probability of a lensing event is often described by the concept of optical depth, which has been well studied in \cite{2018PhRvD..98l3523J} and \cite{2021NatAs...5..560P}. 
The effective lensing cross-section $\sigma$ of one single lens object is defined as
\begin{eqnarray}
\int{ d\sigma}=\int_{y_{\rm min}}^{y_{\rm max}}{\frac{4\pi GM_{\rm Lz}}{c^2} \frac{d_{\rm A}(z_{\rm L},z_{\rm S})}{d_{\rm A}(z_{\rm L})d_{\rm A}(z_{\rm S})} 2y dy}
\label{cross-section}
\end{eqnarray}
where $y_{\rm min}$ and $y_{\rm max}$ are the minimum and maximum impact parameters, respectively. These are determined by the event time of a single peak in QPE and the time delay between the images. 
On the one hand, when the time delay is relatively small (smaller than $y_{\rm min}=f^{-1}(c^3 \Delta t_{\rm min} (1+z_{\rm L})^{-1}(4GM_{\rm Lz})^{-1})$), the two images in QPE cannot be distinguished. 
On the other hand, if $y$ is relatively large (larger than $y_{\rm max}=(1+\varphi_{\rm max}/\varphi_{0})^{1/4}-(1+\varphi_{\rm max}/\varphi_{0})^{-1/4}$, where $\varphi_{\rm max}/\varphi_{0}$ is the peak counts divided by the trigger threshold at that time), the lensing effect becomes very weak. When the flux of the de-magnified image falls below the detection threshold, the lensing effect is considered undetectable. 
Such that the final cross-section of one single lens is
\begin{eqnarray}
\sigma = \frac{4\pi GM_{\rm Lz}}{c^2} \frac{d_{\rm A}(z_{\rm L})d_{\rm A}(z_{\rm L},z_{\rm S})}{d_{\rm A}(z_{\rm S})}(y^2_{\rm max}-y^2_{\rm min}) \Theta (y^2_{\rm max}-y^2_{\rm min})
\label{cross-section_final}
\end{eqnarray}
where $\Theta$ is the Heaviside step function. 
For a lens at redshift $z_{\rm L}$ with number density $n(z_{\rm L})$, the optical depth of a source at redshift $z_{\rm S}$ is defined as
\begin{eqnarray}
\tau = \iiint_{0}^{z_{\rm S}} {n(z_{\rm L}) dV(z_{\rm L})} \int{\frac{d\sigma}{4\pi ^2 d_{\rm A}^2(z_{\rm L})}}
\label{optical_depth1}
\end{eqnarray}
where $n(z_{\rm L})=3H_0^2 \Omega_{\rm L}(1+z_{\rm L})^3(8\pi G M_{\rm Lz})^{-1}$ is the comoving density of lenses with Hubble constant $ H_0=70$ km s $^{-1}$ Mpc $^{-1}$, while $\Omega_{\rm L}$ is the mean lens density. The angular diameter distance $d_{\rm A}(z_{\rm L})$, $d_{\rm A}(z_{\rm S})$ and $d_{\rm A}(z_{\rm L},z_{\rm S})$ can be translated to comoving distance through $\chi(z_{\rm L},z_{\rm S}) = \int^{z_{\rm S}}_{z_{\rm L}}{cdz / H_0 \sqrt{\Omega_{\Lambda} + \Omega_{\rm m} (1+z)^3}}$,  and $\Omega_{\Lambda}=0.714$ and $\Omega_{\rm m} = 0.286$ are the cosmic densities of dark energy and matter, respectively.

\begin{figure}  
\centering
\includegraphics[width=8cm]{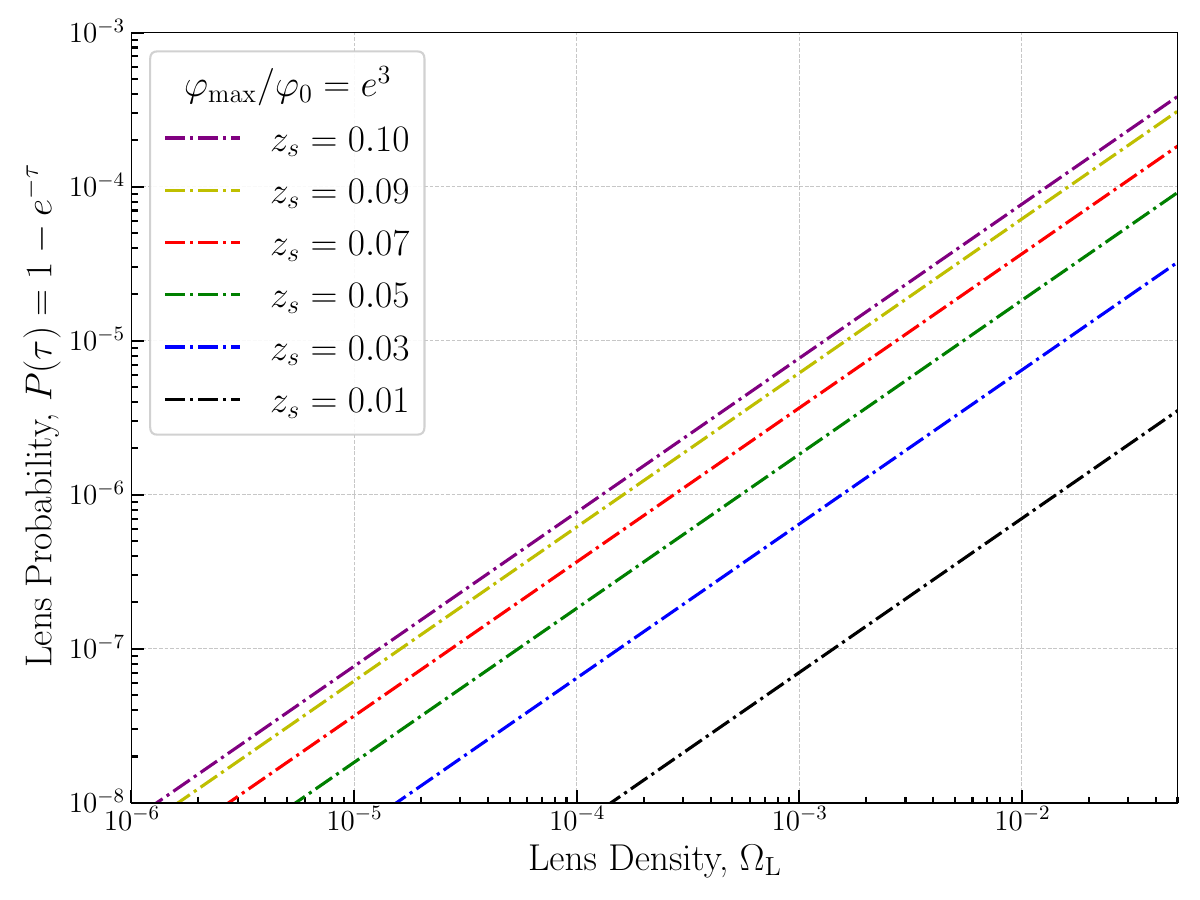}
\caption{Optical depth as a function of lens density $\Omega_{\rm L}$. 
The optical depth is estimated for several source redshifts: $z_{\rm S} = 0.01$ (black dash-dot line), $z_{\rm S} = 0.03$ (blue), $z_{\rm S} = 0.05$ (green), $z_{\rm S} = 0.07$ (red), $z_{\rm S} = 0.09$ (yellow), and $z_{\rm S} = 0.1$ (purple). 
All other parameters are fixed according to the properties of eRO-QPE1 ($z \approx 0.05$): lens mass $M_{\rm Lz} \approx 10^9 M_{\odot}$, minimum time delay $\Delta t_{\rm min} \approx 1.9$ hours, maximum magnification ratio $\varphi_{\rm max}/\varphi_{0} = e^3$, and lens redshift $z_{\rm L} = \frac{1}{2} z_{\rm S}$.}
\label{fig:probability}
\end{figure}

The final form of optical depth in comoving volume is
\begin{eqnarray}
\begin{aligned} 
\tau(\mathbf{x}) &=  \int_0^{z_{\rm S}} \mathrm{d}z_{\rm L} \frac{(1+z_{\rm L})\chi(z_{\rm L})}{\sqrt{\Omega_{\Lambda} + \Omega_{\rm m}(1+z_{\rm L})^3}}\left[\chi(z_{\rm S}) - \chi(z_{\rm L})\right] \quad \\& \times  \frac{3H_0\Omega_{\mathrm{L}}}{2c\chi(z_{\rm S})} \left[y^2_{\max}\left(\varphi_{\text{max}}, \varphi_0\right) - y^2_{\min}\left(\Delta t_{\min}, M_{\rm Lz}, z_{\rm L}\right)\right].
\end{aligned}
\label{optical_depth}
\end{eqnarray}
where $\mathbf{x} \equiv (M_{\rm Lz},z_{\rm L},z_{\rm S},\varphi_{\rm max},\varphi_{0},\Delta t_{\rm min}) $ is all of the parameters that we need to calculate the optical depth. 
Based on the result above, we set $M_{\rm Lz} \approx 10^9 M_{\odot}$, and set the redshift of the source within the range of $[0.01, 0.1]$, while the redshift of the lens is set to half of the source's redshift. 
According to \cite{2022A&A...662A..49A}, a burst is considered to start when the flux reaches $1/e^3$ of the peak flux. Therefore, we set $\varphi_{\rm max}/\varphi_{0}=e^3$. 
The cumulative lens probability as a function of mean lens density are shown in Figure \ref{fig:probability}. 
From the previous formulas, it is evident that $y_{\rm max}$ is related to the quartic root of $\varphi_{\rm max}/\varphi_{0}$, while $y_{\rm min} \propto (\Delta t_{\rm min}/M_{\rm Lz})^{1/2}$.  Therefore, the influence of $y_{\rm max}$ and $y_{\rm min}$ on the probability is relatively limited. The source redshift has a more significant impact on the probability, as indicated in Figure \ref{fig:probability}.

\section{Conclusions and discussion}
\label{sec:summary}
Numerous models have been proposed to explain the origin of QPEs after their discovery. However, most of these models primarily focus on the common characteristics of QPEs, leaving the more intricate behavior of eRO-QPE1 less explored. The complexity seen in eRO-QPE1 raises a crucial question: is this complexity an intrinsic feature of QPEs, or does it point to a separate physical mechanism at play?

In this work, we consider a third possibility: that existing QPE models remain broadly applicable, and the complex structure observed in eRO-QPE1 may arise from gravitational lensing, rather than indicating a fundamentally different physical origin from other known QPE sources.
We find some clues supporting this hypothesis.

We analyze the Obs1 data of eRO-QPE1 and divide it into five energy bands. 
We then obtain the spectral indices in each energy band. 
For each energy band, the complex structure can be separated into two sub-eruptions (via cross-time). 
The two sub-eruptions show similar trends in their narrow-band spectral indices. 
Similarities are also observed in the broad-band hydrogen column density and disk blackbody temperature. 
These consistencies support the interpretation of a common physical origin. 
Therefore, the two sub-eruptions are consistent with being the two images produced by gravitational lensing. 
By fitting with the lens model and using the UltraNest package, we find the time delay of $\Delta t \sim 1.9 \rm h$ and magnification ratio of $\mu^{-1} \sim 0.9$ for the two images, suggesting a lens with mass of $M_{\rm Lz} \sim 10^9 \rm M_\odot$ for both PM and SIS models. 
 
To assess the probability of a lensed QPE, we employ the concept of optical depth, which indicates that such events are expected to be rare. 
If similarly complex light curves are observed in other QPE events like eRO‑QPE1, the lensing interpretation would likely need to be ruled out. 
In that case, the complex variability in obs1 of eRO-QPE1 could then be linked to the intrinsic physical mechanisms of QPEs. 
Consequently, other factors to consider would include the specific mass transfer processes between the companion star and the supermassive black hole \citep{2022ApJ...926..101M,2025MNRAS.543.3503Y}, or alternative physical mechanisms such as the tidal stripping process in a supermassive black hole binary \citep{2025ApJ...994..112D,2025PhRvD.112j3024L}. 
However, it should be noted that the microlensing scenario can also produce similar profiles. 
If we take into account the microlensing scenario, this probability could in fact be one or two orders of magnitude higher \citep{2018MNRAS.474.2975D}. 
We will provide a preliminary discussion of such a scenario in the latter part of this section.

Our calculations find that the source redshift has a more significant impact on the probability, as indicated in Figure \ref{fig:probability}. 
Consequently, eRO-QPE1 may presents the first QPE showing clues of gravitational lensing. 
This study demonstrates that gravitational lensing provides a plausible explanation for the complex structure in eRO-QPE1 obs1  \citep{2021Natur.592..704A}. 
Although our analysis of the static lensing scenario suggests the presence of certain lensing features, some aspects of these interpretations remain open to question. 
We discuss these potential issues below. 

\begin{figure}  
\centering
\includegraphics[width=8.5cm]{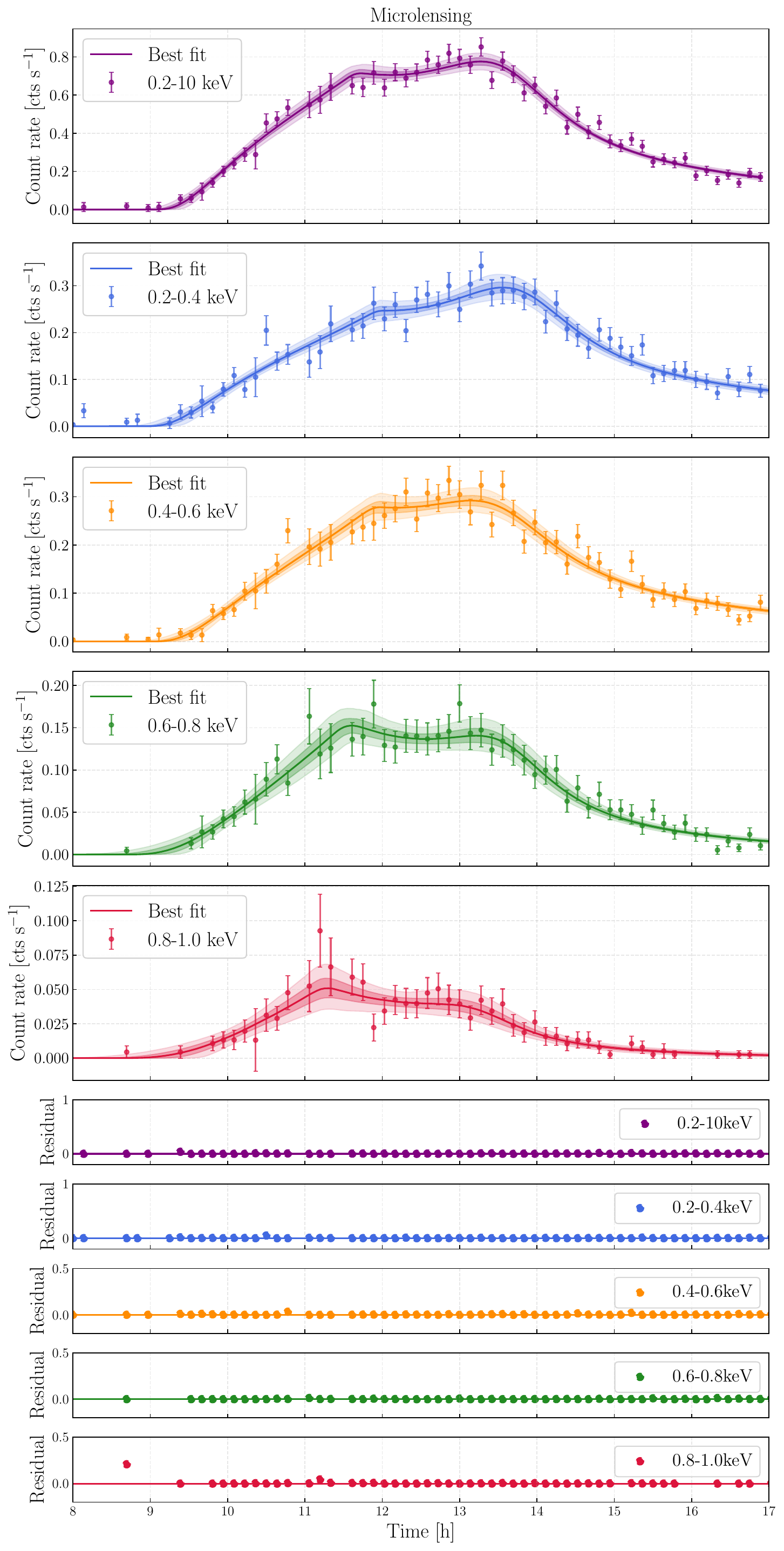}
\caption{Light curve fits under the microlensing scenario: The upper panel shows the fitting results of eRO-QPE1 obs1 data in the energy ranges of $0.2$–$10$ keV (purple), $0.2$–$0.4$ keV (blue), $0.4$–$0.6$ keV (orange), $0.6$–$0.8$ keV (green), and $0.8$–$1.0$ keV (red). The lower panel shows the residual of each data point.} 
\label{fig:MicroLensFit_multi_energy_bands_with_full_and_residuals}
\end{figure}

Among the issues that arise, one natural question is: if a lensing object with mass of $\sim 10^9 M_{\odot}$ are truly at play, then characteristic lensing signatures should manifest in every eruption. 
However, in the observations reported by \citep{2021Natur.592..704A}, such complex structures are not detected in the main peak of eRO-QPE1 Obs2 (one week after eRO-QPE1 Obs1 \citep{2022A&A...662A..49A}). Instead, a relatively clean, single-peaked profile is observed. 
Within the lensing framework, this inconsistency may be reconciled by invoking a dynamic lensing configuration. 
In other words, lensing features only emerge when the lensing object moves into a configuration where the observer, the lens, and the eruption source are momentarily aligned \citep{2024A&A...690L...2K}. 
In this case, the lensing object is not located at an intervening position midway between the observer and the source, and the complex structure observed in the main peak of the light curve at MJD $\sim$ 59057 is no longer caused by the superposition of two images, but rather from microlensing effects produced when the lensing object transits the observer's line of sight toward the eruption source. 
In the following, we discuss this scenario.

In the case of microlensing, the magnification is given by \citep{1996ARA&A..34..419P}

\begin{eqnarray}
\mu = \frac{y^2+2}{y(y^2+4)^{1/2}}
\label{microlensing_magnification}
\end{eqnarray}
where $y=[p^2+((t-t_{\rm max})/t_0)^2]^{1/2}$ is the time-dependent angular distance. $t_{\rm max}$ is the time of highest magnification. The parameter $p$ and $t_0$ can be measured from the light curve. 
The characteristic timescale $t_0$ for a microlensing event is \citep{1996ARA&A..34..419P}, 
\begin{eqnarray}
t_0 = 0.214\,\text{yr} \left( \frac{M_{\rm Lz}}{M_\odot} \right)^{1/2} 
\left( \frac{D_{\rm L}}{10\,\text{kpc}} \right)^{1/2} 
\left(\frac{D_{\rm LS}}{D_{\rm S}} \right)^{1/2} 
\left( \frac{200\,\text{km\,s}^{-1}}{v} \right),
\label{t_0}
\end{eqnarray}
where $v$ is the transverse velocity of the lensing object. 
Using Equations \eqref{eq18} and \eqref{microlensing_magnification}, we performed fits to the light curves in all five energy bands. 
The parameters $A$, $t_{\rm peak}$, $\gamma_1$, $\gamma_2$ were sampled within the same ranges as listed in second colum of Table \ref{tab:best_fit}, where they correspond to the peak flux, time of peak flux, rise timescale, and decay timescale of the initial eruption, respectively. 
The parameter $t_0$ is constrained to be less than 2 hours, while $t_{\rm max}$ is bounded below by a value slightly smaller than the cross-time and above by 15 hours for all bands. 
The parameter $p$—which describes the source position relative to the Einstein radius—is allowed to vary within $[0.01,0.99]$. 

The fits constrain the parameter $p$ to values of approximately $0.55^{+0.03}_{-0.03}$ in the broad 0.2–10 keV band. 
Within the narrower sub-bands, the estimates are approximately: $0.52^{+0.04}_{-0.03}$ (0.2-0.4 keV), $0.59^{+0.05}_{-0.05}$ (0.4-0.6 keV), $0.48^{+0.05}_{-0.04}$ (0.6-0.8 keV), $0.45^{+0.11}_{-0.08}$ (0.8-1.0 keV). 
The fits constrain the parameter $t_{\rm max}$ to values of approximately $13.56^{+0.06}_{-0.06}$ hours in the broad 0.2–10 keV band. 
Within the narrower sub-bands, the estimates are approximately: $13.77^{+0.10}_{-0.10}$ hours (0.2-0.4 keV), $13.52^{+0.11}_{-0.11}$ hours (0.4-0.6 keV), $13.61^{+0.13}_{-0.13}$ hours (0.6-0.8 keV), $13.26^{+0.21}_{-0.21}$ hours (0.8-1.0 keV). 
The fits constrain the parameter $t_{\rm 0}$ to values of approximately $1.63^{+0.02}_{-0.02}$ hours in the broad 0.2–10 keV band. 
Within the narrower sub-bands, the estimates are approximately: $1.71^{+0.04}_{-0.04}$ hours (0.2-0.4 keV), $1.63^{+0.03}_{-0.03}$ hours (0.4-0.6 keV), $1.65^{+0.03}_{-0.03}$ hours (0.6-0.8 keV), $1.56^{+0.05}_{-0.05}$ hours (0.8-1.0 keV). 
The fitting results under the microlensing scenario are shown in Figure \ref{fig:MicroLensFit_multi_energy_bands_with_full_and_residuals}. 
The upper panel shows the fitting results of eRO-QPE1 obs1 data in the energy ranges of $0.2$–$10$ keV (purple), $0.2$–$0.4$ keV (blue), $0.4$–$0.6$ keV (orange), $0.6$–$0.8$ keV (green), and $0.8$–$1.0$ keV (red). 
The lower panel shows the residual of each data point, which are same as Figure \ref{fig:total_fitted_image}. 
The shaded area represents their uncertainty. 

Based on Equation. \eqref{t_0}, we proceed to discuss the fitting results under two approximations. 
The first approximation is $D_{\rm LS} \approx D_{\rm S}$, in which Equation. \eqref{t_0} reduces to $t_0 = 0.214\,\text{yr} (M_{\rm Lz}/M_\odot)^{1/2} (D_{\rm L}/10\,\text{kpc})^{1/2} (200\,\text{km/s}^{-1}/v).$ 
Such that 
\begin{align}
M_{\rm Lz} \approx M_{\odot} (\frac{t_0}{0.214 \ \rm{yr}})^{2} (\frac{D_{\rm L}}{10 \rm{kpc}})^{-1} (\frac{v}{200 \rm{km/s}})^{2}
\label{ML_near}
\end{align}  
Substituting $t_0$ into Equation \eqref{ML_near} with $D_{\rm L}$ between 1 and 10 kpc yields a lens mass of approximately $10^{-5}$ to $10^{-6}M_{\odot}$. 
In the second approximation, $D_{\rm L} \approx D_{\rm S}$, Equation. \eqref{t_0} reduces to $t_0 = 0.214 \ \text{yr} \ (M_{\rm Lz}/M_\odot)^{1/2} (D_{\rm LS}/10 \ \text{kpc})^{1/2} (200 \ \text{km s}^{-1}/v)$, where $v=\sqrt{GM_{\rm SMBH}/D_{\rm LS}}$ and $M_{\rm SMBH}$ is the central SMBH mass of the host galaxy. 
In this case, 
\begin{align}
M_{\rm Lz} \approx M_{\odot} (\frac{t_0}{0.214 \ \rm{yr}})^{2} (\frac{D_{\rm LS}}{10 \rm{kpc}})^{-1} (\frac{v}{200 \rm{km/s}})^{2}.
\label{ML_local}
\end{align}  
In this case, $t_0$ is jointly determined by the mass of the central supermassive black hole ($M_{\rm SMBH}\sim 10^6M_{\odot}$) in the host galaxy of the QPE, $D_{\rm LS}$, and $M_{\rm Lz}$. 
With $t_0$ substituted into Equation. \eqref{ML_local} and $D_{\rm LS}$ set to the range $10^3R_{\rm S}(M_{\rm SMBH}) \lesssim D_{\rm LS} \lesssim 10^5R_{\rm S}(M_{\rm SMBH})$, the resulting lens mass is on the order of a few to $10^4 M_{\odot}$. 

Building on these estimates, the microlensing scenario indeed yields a more plausible lens mass range than strong lensing models, but it also brings its own set of challenges. 
In particular, the detailed physics of QPE formation and the complex environmental conditions surrounding these events remain uncertain. 
We intend to address these intricacies and present a more comprehensive microlensing analysis in future work.

\section*{DATA AVAILABILITY}
The data underlying this article will be shared on reasonable request to the corresponding author.

\section*{Acknowledgements}
We thank Arcodia for providing the light curve data, which were essential for our light curve analysis. 
We thank Meng-Ye Wang for helpful discussions. This work is supported by the National Key R\&D Program of China (No. 2023YFC2205901), the National Natural Science Foundation of China under grants 12473012 and 12533005.

\bibliographystyle{mnras}
\bibliography{ref} 

\end{document}